\documentclass[11pt,a4paper]{article}
\pdfoutput=1
\usepackage{jheppub}
\usepackage{amsthm}

\usepackage{hyperref}

\usepackage{physics}
\usepackage{xcolor}
\usepackage{mathtools}
\usepackage{tensor}
\usepackage{empheq}

\usepackage{tikz}
\usepackage{graphicx}
\usepackage{subfigure}
\usetikzlibrary{calc,fadings,decorations.pathreplacing}
\usepackage{verbatim}

\usepackage{url}

\DeclareMathOperator{\MyProd}{\scalebox{1.4}{$\mathrm{I\kern-0.2ex I}$}}

\tikzset{%
  >=latex, 
  inner sep=0pt,%
  outer sep=2pt,%
  mark coordinate/.style={inner sep=0pt,outer sep=0pt,minimum size=3pt,
    fill=black,circle}%
}

\preprint{LCTP-22-11}

\title{Cardy Expansion of 3d Superconformal Indices\\
and Corrections to the Dual Black Hole Entropy }

\author[a]{\small{Alfredo Gonz\'alez Lezcano}}

\author[b]{\small{Maximilian Jerdee}}

\author[b,c,d]{\small{Leopoldo A. Pando Zayas}}

\emailAdd{alfredo.gonzalez@apctp.org, mjerdee@umich.edu, lpandoz@umich.edu}

\affiliation[a]{Asia Pacific Center for Theoretical Physics  
 Postech, Pohang 37673, South Korea}

\affiliation[b]{Leinweber Center for Theoretical Physics 
University of Michigan, Ann Arbor, MI 48109, USA}
\affiliation[c]{School of Natural Sciences, Institute for Advanced Study, Princeton, NJ 08540, USA}

\affiliation[d]{The Abdus Salam International Centre for Theoretical Physics, 34014 Trieste, Italy}

\abstract{We consider the superconformal index of three-dimensional ${\cal N}=2$ supersymmetric field theories computed via localization on $S^1\times S^2$. We systematically develop an expansion where the ratio of the radius of $S^1$ to the radius of $S^2$ is taken very small -- a Cardy-like expansion. We emphasize the sub-leading structures in this Cardy-like expansion as well as their interplay with the large-$N$ limit for theories with gauge group of the form product of $U(N)$ factors. We demonstrate that taking the large-$N$ limit first leads to an expression for the index that only includes terms proportional to  $1/\beta$ and powers of  $\beta^{i=0,1,2}$ where $\beta$ is the ratio of radii. As we depart from the $\beta\to 0$ limit for finite $N$, we find indications of non-perturbative contributions  of the form $e^{-1/\beta}$. For the ABJM theory we explore the implications of the Cardy-like expansion for corrections to the entropy of the rotating, electricallly charged, asymptotically AdS$_4$ dual black hole. Interestingly, we find that the corrections in $\beta$ to the entropy can be accounted for by appropriately shifting the electric charges and the angular momentum.}


\arxivnumber{}

\begin{document}

\maketitle
\today

\newpage

\section{Introduction}
The entropy of rotating, electrically charged, asymptotically AdS$_4$ black holes has recently been provided a microscopic foundation via the three-dimensional superconformal index computed using supersymmetric localization on $S^1\times S^2$ \cite{Choi:2019zpz,Nian:2019pxj}.  The analysis extended groundbreaking work for asymptotically AdS$_5$ black holes via the four-dimensional superconformal index (SCI) \cite{Cabo-Bizet:2018ehj,Choi:2018hmj,Benini:2018ywd}. These developments have provided a  window into the black hole entropy of rotating, electrically charged, asymptotically AdS  black holes via the AdS/CFT correspondence. Remarkably, there is now potential access to the full quantum entropy of AdS black holes, that is, to the leading Bekenstein-Hawking part as well as to sub-leading contributions. For example, a series of works successfully addressed logarithmic entropy corrections to  AdS$_4$ black holes  \cite{Liu:2017vbl,Gang:2019uay, Benini:2019dyp,PandoZayas:2020iqr}. Higher curvature corrections for AdS$_4$ have also been considered in \cite{Bobev:2020egg,Bobev:2020zov, Bobev:2021oku,Ghosh:2020rwf} and, more recently, for AdS$_5$ black holes \cite{Liu:2022sew, Bobev:2022bjm,Cassani:2022lrk}.

For the 4d SCI, the Cardy-like expansion was systematically developed in \cite{GonzalezLezcano:2020yeb,Lezcano:2021qbj} where it was further shown that the full index truncates as an expansion in small $\beta$. The logarithmic corrections to the SCI were elucidated in \cite{GonzalezLezcano:2020yeb}  (see also  \cite{Amariti:2020jyx,Amariti:2021ubd}). The truncation of the 4d SCI motivated later work to re-derive these results in the natural language of  effective field theory in  \cite{Cassani:2021fyv,ArabiArdehali:2021nsx}.  Motivated by these improvements in our understanding of the 4d SCI, we turn to explorations of the 3d SCI in the Cardy-like expansion.

Another important motivation for studying the SCI beyond the strict Cardy limit  comes from the hope of better understanding the role of the Cardy-like expansion and the Kerr/CFT$_2$ correspondence. The Cardy limit seems to be a ubiquitous ingredient in the derivation of the leading term in the SCI in various dimensions. An interpretation from the gravitational point of view, advanced in \cite{David:2020ems},  suggests that the Cardy limit might be related to taking the near-horizon limit in the rotating black hole geometry along the lines of Bardeen and Horowitz \cite{Bardeen:1999px}. This potential relation raises the natural question of how much of the full black hole entropy can be recovered in this near-horizon limit using the Kerr/CFT$_2$ correspondence \cite{Guica:2008mu,Lu:2008jk}. Modulo global issues materialized through zero modes, the authors of \cite{David:2021qaa} showed that one can track logarithmic corrections to the entropy of asymptotically AdS$_5$ black holes and black strings. In asymptotically  AdS$_4$ spacetimes, however,  taking the near-horizon limit affects the nature of zero modes. Nevertheless, by tracking the degrees of freedom as one takes the near-horizon limit, the authors of \cite{David:2021eoq} showed that the local degrees of freedom lead to the same contribution and identified that the culprit are the zero mode contributions. 

We, therefore, study the Cardy-like expansion of the 3d ${\cal N}=2$ SCI in general and discuss the ABJM case in full detail. We also discuss the implications of the $\beta$-corrections to the index for the entropy of the dual AdS$_4$ black hole. 

The manuscript  is organized as follows. We review the superconformal index of 3d ${\cal N}=2$ supersymmetric field theories in section \ref{Sec:SCI}; we include both a Hamiltonian and a Lagrangian description with its subsequent localization approach. Section \ref{Sec:ABJM} develops the Cardy-like expansion in the explicit example of the ABJM theory. In section \ref{Sec:LargeN} we show that when first expanded in the large-$N$ limit, the index truncates as an expansion in small $\beta$ for a large class of theories. We explore the implications for the entropy of the dual black holes in section \ref{Sec:Entropy}. We conclude in section \ref{Sec:Conclsions} and relegate a number of technical details to a series of appendices. 

\section{The Superconformal Index of $\mathcal{N}=2$ Supersymmetric theory on $ S^1\times S^2$}\label{Sec:SCI}

Consider an $\mathcal{N} = 2$ SCFT 
defined on $ S^1 \times S^2$ with gauge group $G$. Let the radius of $S^2$ be set to $1$ for simplicity and $\beta^{\prime}+ \beta$ the period of $S^1$. The bosonic part of the 3d $\mathcal{N}=2$ superalgebra is $SO(2,3) \times SO(2)$. One can naturally associate to the Cartan generators of the superconformal algebra the charges $\epsilon, j_3$ and $R$ living in the factors: $SO(2)_{\epsilon} \times SO(3)_{j_3} \times SO(2)_R  \subset SO(2,3) \times SO(2) $. Let us denote the  supersymmetry generators as $\mathcal{Q}$ and the special supercharges, which are Hermitian conjugate of the $\mathcal{Q}'s$, as  $\mathcal{S}$: $\mathcal{S} = \mathcal{Q}^{\dagger}$. It is then possible to define a quantity called the superconformal index, defined  originally for 4d theories in \cite{Romelsberger:2005eg, Kinney:2005ej} and subsequently extended to  dimensions 3, 5, and 6 in \cite{Bhattacharya:2008zy}, the relevance of certain magnetic monopole contributions in the three-dimensional case was discussed in \cite{Kim:2009wb}  . The SCI is a function of the spectrum that is forced by superconformal symmetry to remain unchanged under continuous deformations of parameters controlling the spectrum.
Selecting a special pair of such charges, we can define the superconformal index \cite{Bhattacharya:2008zy}:
\begin{align}
\mathcal{I}_{S^2}(\beta,\xi,n) & = \text{Tr}_{\mathcal{H}_{\text{BPS}(S^2;n)}} \left[(-1)^F e^{- \beta^{\prime} \{\mathcal{Q},\mathcal{S}\} - \beta \left(\epsilon +j_3\right) -\xi m}\right]. \label{eq:Ind0}
\end{align}
The trace runs over the Hilbert space, $\mathcal{H}_{BPS}(S^2;n)$,  of BPS states in Hamiltonian quantization on $S^2$ with $m$ denoting the magnetic flux on $S^2$. Here $n$ denotes the external magnetic fluxes of the flavor symmetry. The combinations $\epsilon + j_3$ and $\{\mathcal{Q}, \mathcal{S}\} = \epsilon - R- j_3$ commute with both $\mathcal{Q}$ and $\mathcal{S}$.  As usual, we denote as $F$ the fermion number and  introduce $e^{- \xi m}$ to account for magnetic fluxes on $S^2$ associated to the $U(1)$ factors of the gauge group. Should there be additional conserved charges commuting with the supercharges one can turn on chemical potentials associated to them and insert the corresponding fugacities in \eqref{eq:Ind0}:

\begin{align}
\mathcal{I}_{S^2}\left(q,z_l, y,n\right)& = \text{Tr}_{\mathcal{H}_{\text{BPS}(S^2;n)}} \left[\left(-1\right)^F e^{- \beta^{\prime}\{\mathcal{Q}, \mathcal{S}\}- \beta \left(\epsilon+j_3\right)}e^{-i \gamma_l h_l - \xi m}\right], \label{eq:3dIndex}
\end{align}
where $q=e^{-\beta}, \hspace{2mm} y= e^{-\xi}$ and $z_{l}=e^{-i\gamma_{l}}$ ($l =1, \cdots, \#- \text{of flavors}$). The distinguished supercharges $\mathcal{Q}$ and $\mathcal{S}$ satisfy the relations:
\begin{align}
\mathcal{Q}^2 =\mathcal{S}^2 =0, \hspace{2mm}\{\mathcal{Q}, \mathcal{S} \} = \epsilon- R- j_3.
\end{align}
Namely, $\mathcal{Q}$ and $\mathcal{S}$ are nilpotent and the anticommutator equation implies the BPS energy bound $\epsilon \geq R +j_3$. The index receives contributions from states saturating the BPS energy bound and, therefore, it is independent of $\beta^\prime$, it can be rewritten simply as
\begin{align}
\mathcal{I}_{S^2}(q,f_k, n) = \text{Tr}_{\mathcal{H}_{\text{BPS}(S^2;n)}}  \left[(-1)^F e^{-\beta (R + 2 j_3)} f_k^{\Phi_k}\right], \label{eq:indexshort}
\end{align}
where $f_k =e^{- M_k}$ and we have used a collective set of chemical potentials $M_k=\{i\gamma_l, \xi\}$ and the charges $\Phi_k=\{h_l, m\}$. 


\subsection*{The $R$-charge assignment and the superconformal index as a function of $\beta$}

The standard choice for the fermion number is $F= 2 j_3$, however another useful choice is $F=R$ which gives rise to the so-called ``$R$-charge index'', studied in \cite{Choi:2019dfu} (see also \cite{Cassani:2021fyv} for the 4d ``$R$-charge index''). 
A crucial observation in \cite{Choi:2019dfu} is to use the integer valued trial $R$-charges, which guaranties that, upon shifting $\beta \rightarrow \beta + i \pi$ the $(-1)^F$ insertion is naturally generated by $e^{i \pi R}$, as opposite to a non-trivial phase that a non-integer $R$-charge would produce. 
 By appropriately tuning the parameter $\beta$ one can produce extra phases inside the trace of \eqref{eq:indexshort}, and hence modify the interplay between different contributions to the total sum. To illustrate this phenomenon and its implications, we analyze how a generic shift of the form $\beta \rightarrow \beta + 2 \pi i \kappa$ ($\kappa \in \mathbb{R}$) affects $\mathcal{I}_{S^2}(q, f_k, n)$:

\begin{align}
\mathcal{I}_{S^2}(q,f_k, n) = \text{Tr}_{\mathcal{H}_{\text{BPS}(S^2;n)}}  \left[(-1)^F e^{- 2 \pi i \kappa(R+ 2 j_3)} e^{-\beta (R + 2 j_3)}f_k^{\Phi_k}\right] \label{eq:indexshift}
\end{align}
Note that the same effect would have been produced if in the definition of the index \eqref{eq:3dIndex} we shifted  $\beta^{\prime} \rightarrow \beta^{\prime} - 2 \pi i \kappa$ as well. 
For $\kappa = \frac{1}{2}$ one obtains the shift that implements the transition between the $R$-charge index and the standard index, which can be made explicit by defining $F_r\equiv R$ and $F_j \equiv 2 j_3$, therefore:
\begin{align}
\mathcal{I}_{S^2}(q,f_k, n) = \text{Tr}_{\mathcal{H}_{\text{BPS}(S^2;n)}}  \left[(-1)^F (-1)^{F_r}(-1)^{F_j} e^{-\beta (R + 2 j_3)}f_k^{\Phi_k}\right]. \label{eq:indexshiftRj3}
\end{align}
It is  now clear that if we start with $F=F_j$, the shift with $\kappa = \frac{1}{2}$ yields the $R$-charge index and $\kappa$ interpolates between the two indices when varied within the interval $[0, \frac{1}{2}]$. 
 
The authors of \cite{Cassani:2021fyv, ArabiArdehali:2021nsx} adopted a more systematic and instructive point of view when analyzing the 4d superconformal index on $S^1 \times S^3$ . Let us denote the superconformal index in 4d as $\mathcal{I}^{\text{4d}}(\omega_1, \omega_2)$, where $\omega_{1,2}$ are the angular velocities on $S^3$. In this context we can learn important lessons by considering $\mathcal{I}^{\text{4d}}(\omega_1, \omega_2)$ as a complex function of its arguments. 
 
 As emphasized in \cite{Cassani:2021fyv}, since the $R$-charges of some fields in new-minimal supergravity are not integral, the multi-valuedness of $\mathcal{I}^{4d}(\omega_1, \omega_2)$ arises because it is necessary
to choose an $R$-symmetry gauge field, and if the $R$-charges of various fields are not
integral, certain large gauge transformations are not allowed. For instance, in $\mathcal{N}=4$ super Yang-Mills theory, all bosonic elementary fields have $R$-charges of the form $\frac{2 \mathbb{Z}}{3}$, all the fermionic fields have $R$-charges of the form $\frac{2 \mathbb{Z}+1}{3}$. As a result,  only shifting $\omega_1 \rightarrow \omega_1 \pm 6 \pi i$ or $\omega_2 \rightarrow \omega_2 \pm 6 \pi i$ leaves $\mathcal{I}^{4d}(\omega_1, \omega_2)$ invariant.  In addition it is possible to perform $\omega_{1,2} \rightarrow \omega_{1,2} \pm 4 \pi i$ and still leave $\mathcal{I}^{4d}(\omega_1, \omega_2)$ invariant. This freedom can be exploited to set $\text{Im}(\omega_1) \in [0, 6 \pi)$ and $\text{Im}(\omega_2) \in [0, 2 \pi)$. We then conclude that $\mathcal{I}^{4d}(\omega_1, \omega_2)$ takes values on a triple cover of the space of complex fugacities $\omega_{1,2}$ space.

It is natural to question whether a similar kind of multi-valuedness arises for $\mathcal{I}_{S^2}(q,f_k, n)$ if we select the exact IR $R$-charge. Let us then study how the evaluation of $\mathcal{I}_{S^2}(q,f_k,n)$ using supersymmetric localization is carried with a shifted $\beta$.

\paragraph{Shifting chemical potentials} \par
 When we study the specific example of the SCI of ABJM in section \ref{Sec:ABJM}, we will see it convenient to shift the chemical potentials to evaluate the index. This motivates us to study what effect would such shifting have at the level of the Hamiltonian definition of the SCI. 
Let us ignore the magnetic flux for the moment and consider now the effect of shifting the chemical potentials $\gamma_l \rightarrow \gamma_l - \frac{i }{2} \beta$ in \eqref{eq:indexshort}:
\begin{align}
    \mathcal{I}_{S^2}(q,f_k, n) &= \text{Tr}_{\mathcal{H}_{\text{BPS}(S^2;n)}}  \left[(-1)^F e^{-\beta (R + 2 j_3)}e^{i (\gamma_l- \frac{i }{2} \beta) h_l}\right] \label{eq:indexFshift} \\
    & = \text{Tr}_{\mathcal{H}_{\text{BPS}(S^2;n)}}  \left[(-1)^F e^{-\beta (R - \frac{1}{2}\sum_{l}h_l+ 2 j_3)}e^{i \gamma_l h_l}\right]
\end{align}
\par 
Note now that appropriate changes in the $R$-charge can affect the value of $(-1)^{F_r}$, hence we can use the following redefinition $R = R^{\prime} + \frac{1}{2}\sum_l h_l$, therefore we have:
\begin{align}
 \mathcal{I}_{S^2}(q,f_k, n) &=   \text{Tr}_{\mathcal{H}_{\text{BPS}(S^2;n)}}  \left[(-1)^F e^{-\beta (R^{\prime} + 2 j_3)}e^{i \gamma_l h_l}\right] 
\end{align}
Here we can see that $R^{\prime}$ may take non- integer values, which would transform $(-1)^{F_r}$ into a generic phase. 

Another type of shifting is the one that interpolates between the two types of indices, namely the $R$-charge index and the standard index. When evaluating the index in the Cardy-like expansion, we first choose what index we are evaluating and then implement the expansion for small values of the given $\beta$. In section \ref{Sec:Entropy} we will be interested in extracting the dual black holes entropy via a Laplace transform of the SCI that takes us from the grand-canonical ensemble to the micro-canonical one. This evaluation is done in the saddle point approximation and singles out critical values of the chemical potentials that dominate the integration. We will explicitly see that, even if we perform the expansion assuming small values of $\beta$, there is a regime of the charges of the dual black hole, where the critical value approaches the shifted $\beta$. We may see this as an indication that such regime of charges is more suitably probed using a Cardy-like expansion of the $R$-charge index.


\subsection{Localization results}
We now proceed to review the evaluation of $\mathcal{I}_{S^2}(q,f_k, n)$ using supersymmetric localization. The starting point is to consider the index as a partition function defined via a path integral that schematically has the form:
\begin{align}
    Z & = \int \mathcal{D} \Psi e^{- S[\Psi]}, \label{eq:PI}
\end{align}
where $S[\Psi]$ is the Euclidean action of the $\mathcal{N}=2$ theory and $\Psi$ is a generic set of fields of the theory. The evaluation of $\mathcal{I}_{S^2}$ for arbitrary $R$-charge assignment has been performed in \cite{Imamura:2011wg} and we closely follow their development in our presentation. We shall pay close attention to the boundary conditions imposed on the fields $\Psi$ depending of whether we probe the $R$-charge index or the standard superconformal one.  The usual localization argument requires the selection of a supersymmetric transformation $\delta$ compatible with the background in which the theory is defined and a deformation of the action by a  $\delta$-exact term is then introduced $S[\Psi] \rightarrow S[\Psi] +  t \delta \mathcal{V}$ with $t \in \mathbb{R}$ yielding:
\begin{align}
  Z & = \int \mathcal{D} \Psi e^{- S[\Psi] - t \delta \mathcal{V}}. \label{eq:PI1}   
\end{align}

With a suitable choice of $\mathcal{V}$,  and noticing that $Z$ is independent of $t$, it is possible to render  Gaussian the path integral for all but a finite number of degrees of freedom in the $t \rightarrow \infty$ limit.
 The supersymmetric transformation $\delta$ needed to deform the action is parametrized by some Killing spinors. In general, there are $8$ supercharges in the $\mathcal{N}=2$ superconformal algebra, four parametrized by a Killing spinor $\epsilon$ and the other four by the Killing spinor $\Bar{\epsilon}$ both of which satisfy  Killing spinor equations of the from:
    \begin{align}
        D_{\mu} \epsilon & = \gamma_{\mu} \zeta, \hspace{3mm} D_{\mu} \Bar{\epsilon} = \gamma_{\mu} \Bar{\zeta}
    \end{align}
    for arbitrary spinors $\zeta$ and $\Bar{\zeta}$. Here $D_{\mu}$ is the covariant derivative and $\gamma_{\mu}$ are the Dirac Gamma matrices. The supersymmetric transformation parametrized by the Killing spinor $\epsilon$ is denoted as $\delta_{\epsilon}$.
    
    Let us now study how the different multiplets transforms under supersymmetry.
    For a vector multiplet $(A_{\mu}, \sigma, D, \lambda)$ the transformations under $\delta_{\bar{\epsilon}}$ have the form:
    \begin{align}
        \begin{split}
            \delta_{\Bar{\epsilon}}A_{\mu} & =- i \Bar{\epsilon}\gamma_{\mu}\lambda, \hspace{3mm}\delta_{\Bar{\epsilon}}\sigma = \Bar{\epsilon} \lambda, \hspace{3mm} \delta_{\Bar{\epsilon}} \lambda = 0, \\
            \delta_{\Bar{\epsilon}} D & = i \Bar{\epsilon}\gamma^{\mu}D_{\mu}\lambda + i \Bar{\epsilon}[\sigma, \lambda] + \frac{i}{3}D_{\mu}\Bar{\epsilon}\gamma^{\mu} \lambda, \\
            \delta_{\Bar{\epsilon}}\Bar{\lambda} & = - \frac{i}{2}\gamma^{\mu \nu} \bar{\epsilon}F_{\mu \nu} -\gamma^{\mu}\Bar{\epsilon} D_{\mu} \sigma + i D \Bar{\epsilon} - \frac{2}{3}\gamma^{\mu}D_{\mu}\Bar{\epsilon}\sigma.
        \end{split}
    \end{align}
    For a chiral multiplet $(\phi, \psi, F)$ with $R$-charge $r_{\phi}$ the supersymmetric transformation is:
     \begin{align}
         \begin{split}
                 \delta_{\Bar{\epsilon}}\phi^{\dagger} & = \sqrt{2} \Bar{\epsilon} \Bar{\psi}, \hspace{3mm}   \delta_{\Bar{\epsilon}} \phi = 0, \hspace{3mm}   \delta_{\Bar{\epsilon}} \Bar{\psi} = \sqrt{2}i \Bar{\epsilon} F^{\dagger}, \hspace{3mm}   \delta_{\Bar{\epsilon}}F^{\dagger} = 0, \\
                   \delta_{\Bar{\epsilon}} \psi &= \sqrt{2} \Bar{\epsilon} \sigma \phi - \sqrt{2} \gamma^{\mu} \Bar{\epsilon}D_{\mu}\phi - \frac{2 \sqrt{2}}{3}r_{\phi} \phi \gamma^{\mu}D_{\mu}\Bar{\epsilon}, \\
                     \delta_{\Bar{\epsilon}} F & = \sqrt{2}i \Bar{\epsilon}\gamma^{\mu}D_{\mu}\psi + \sqrt{2}i \Bar{\epsilon}\sigma \psi +2 i \Bar{\epsilon}\Bar{\lambda} \phi + \frac{2 \sqrt{2}}{3}\left(r_{\phi}- \frac{1}{2}\right)D_{\mu}\Bar{\epsilon}\gamma^{\mu}\psi.
         \end{split}
     \end{align}
     The deformation to the action receives contribution from both  the chiral multiplet and the vector multiplet, therefore we split $\delta \mathcal{V} = \delta \mathcal{V}^{\text{gauge}} + \delta \mathcal{V}^{\text{chiral}}$. Choosing two linearly independent Killing spinors $\Bar{\epsilon}_1$ and $\bar{\epsilon}_2$ we can then write:
     \begin{align}
         \begin{split}
               \delta_{\Bar{\epsilon}_1} \mathcal{V}^{\text{gauge}} & =   \delta_{\Bar{\epsilon}_1}  \delta_{\Bar{\epsilon}_2} \left( \int \sqrt{g} d^3 x \text{Tr} \left(- \frac{1}{2} \Bar{\lambda} \Bar{\lambda}\right)  \right.\\
               &\left. + \int \sqrt{g}d^3 x \text{Tr}\left[V_{\mu}V^{\mu} + D^2 - 2 \Bar{\lambda} \gamma^{\mu}D_{\mu}\lambda - 2 \Bar{\lambda}[\sigma, \lambda] - \Bar{\lambda} \gamma_3 \lambda\right]\right) \\
                 \delta_{\Bar{\epsilon}_1} \mathcal{V}^{\text{chiral}} & =   \delta_{\Bar{\epsilon}_1}  \delta_{\Bar{\epsilon}_2} \left(\int \sqrt{g}d^3 x \left(- \frac{1}{2} \phi^{\dagger}F\right)\right.\\
                 & \left.+ \int \sqrt{g}d^3x \left[(1- 2 r_{\phi})\left(\phi^{\dagger}D_3\phi + \frac{1}{2}\bar{\psi} \gamma_3 \psi\right) + r_{\phi}(1- r_{\phi})\phi^{\dagger} \phi\right] \right)
         \end{split}
     \end{align}
     where $\int \sqrt{g}d^3x$ denotes the integration measure over $S^2\times S^1$ and the vector $V_{\mu}$ is defined as:
     \begin{align}
         V_1 & = F_{23} - D_1 \sigma, \hspace{3mm} V_2 = F_{31} - D_2 \sigma, \hspace{3mm} V_3 = F_{12}- D_3 \sigma- \sigma.
     \end{align}
     Reducing the partition function $Z$ defined through the path integral \eqref{eq:PI1} to the superconformal index on $S^2 \times S^1$ requires imposing twisted boundary conditions on the fields. We want to ensure that all fields in our theory fulfill boundary conditions which are compatible with supersymmetry. To this end, let us consider how the Killing spinor behaves along the compact Euclidean time direction $\tau$:
     \begin{align}
         \bar{\epsilon}_1(\tau + \beta^{\prime} + \beta) & = e^{\frac{\beta^{\prime} +\beta}{2}} \bar{\epsilon}_1(\tau), \label{eq:KillingBC}
     \end{align}
     therefore, if we assign the following quantum numbers of $\bar{\epsilon}_1$:
     \begin{align}
         R(\bar{\epsilon}_1) = -1, \hspace{3mm} j_3(\bar{\epsilon}_1) = \frac{1}{2}, \hspace{3mm} F_l(\bar{\epsilon}_1) =0.
     \end{align}
we have:
\begin{align}
     \bar{\epsilon}_1(\tau + \beta^{\prime} + \beta) & = e^{-(R+ j_3)\beta^{\prime} +  j_3 \beta+ F_l \gamma_l} \bar{\epsilon}_1(\tau).
\end{align}
Therefore we impose 
\begin{align}
    \Psi(\tau + \beta^{\prime} + \beta) & = e^{-(R+ j_3)\beta^{\prime} +  j_3 \beta+ F_l \gamma_l}  \Psi(\tau ) \label{eq:BC}.
\end{align}
Note that, shifting $\beta \rightarrow \beta + 2 \pi i \kappa$ and $\beta^{\prime} \rightarrow \beta^{\prime} - 2 \pi i \kappa$, leaves \eqref{eq:KillingBC} invariant, the boundary condition \eqref{eq:BC} is modified in the following way:
\begin{align}
   \Psi(\tau + \beta^{\prime} + \beta) & = e^{-(R+ j_3)\beta^{\prime} +  j_3 \beta+ F_l \gamma_l + 2 \pi i \kappa (2j_3 + R)}  \Psi(\tau ) \label{eq:BCmod} , 
\end{align}
 Notably, the path integral localizes around $V_{\mu}=0$ \cite{Imamura:2011wg} which has the following solutions:
      \begin{align}
          A_{\mu}^{\text{saddle}} & = \frac{u}{\beta^{\prime}+ \beta} d \tau + \mathfrak{m} B_i d \theta^i, \hspace{3mm} \sigma^{\text{saddle}} = \frac{\mathfrak{m}}{2} \label{eq:locsad}
      \end{align}

where $\tau$ is the coordinate along $S^1$ and $\theta_{i}, \hspace{2mm} i=1,2$ are the angular coordinates parametrizing $S^2$. We denote with $u$ the Wilson line along $S^1$ which takes values in the Cartan of the Lie algebra of the gauge group $G$. The magnetic charge of the Dirac monopole configuration is $\mathfrak{m}$ and it is also an element of the Cartan subalgebra of $G$. With $B_i$ we denote the Dirac monopole with unit magnetic charge.

Turning on canonically normalized fluctuations $\widetilde{\Psi}$ around the localization saddle \eqref{eq:locsad} we have:
\begin{align}
    \Psi = \Psi^{\text{saddle}} + \frac{1}{\sqrt{t}} \widetilde{\Psi} \label{eq:fluctuation}.
\end{align}
Upon replacing \eqref{eq:fluctuation} into \eqref{eq:PI1} with the deformed action and taking the $t \rightarrow \infty$ limit the path integral reduces to:
\begin{align}
    \mathcal{I}_{S^2} & = \frac{1}{|W_G|} \sum_{\mathfrak{m} \in \Gamma_G^\vee} \int \left(\prod_{a=1}^{\text{rk}(G)}\frac{d u_a}{2\pi i} \right)\int \mathcal{D}\widetilde{\Psi} e^{- S^{\text{class}}} e^{- \int \sqrt{g} d^3 x \widetilde{\Psi} \mathcal{O}\widetilde{\Psi}}, \label{eq:intLocal1}
\end{align}
where $S^{\text{class}}= S[\Psi^{\text{saddle}}] $ and  $\mathcal{O}$ is the appropriate differential operator and $\mathfrak{m}$ belongs to the dual root lattice $\Gamma^\vee_G$ including Weyl equivalent roots, hence the order of the Weyl group $|W_G|$ in the denominator. After considering the individual contributions coming from different fields $\Psi$, then \eqref{eq:intLocal1} can be written as:
\begin{align}\label{Eq:3dSCI}
    \mathcal{I}_{S^2} & =  \frac{1}{|W_G|} \sum_{\mathfrak{m} \in \Gamma_G^\vee} y^{m}  \int \left(\prod_{a=1}^{\text{rk}(G)}\frac{d u_a}{2\pi i} \right) e^{- S_{\text{CS}}^{\text{class}}} Z^{\text{gauge}}_{1\text{-loop}} Z_{1\text{-loop}}^{\text{chiral}}.
\end{align}

The contribution $Z_{1\text{-loop}}^{\text{chiral}}$ is given as follows:

\begin{align}
   Z_{1\text{-loop}}^{\text{chiral}} = \frac{\det D_{\psi}}{\det D_{\phi}},
\end{align}
where $D_{\psi}$ and $D_{\phi}$ are the differential operators associated to the respective kinetic terms of the fermion and complex scalar of the chiral multiplet. Let us focus first on $D_{\phi}$:
\begin{align}
    D_{\phi} = - D_3 D_3 - D_i D_i - s^2 + r_{\phi}(1- r_{\phi}) + (1- 2 r_{\phi})D_3. 
\end{align}
The eigenvalues of $D_{\phi}$ are given by:
\begin{align}
    D_{\phi}  = (j + r_{\phi} + \Delta_3)(j+1 - r_{\phi} - \Delta_3), \label{eq:Dphi}
\end{align}
where $\Delta_3$ is the eigenvalue of $D_3$ which has the form:
\begin{align}
    \Delta_3 = \frac{1}{\beta^{\prime}+\beta} \left[2 \pi i n - i w(u) - (R+ j_3)\beta^{\prime} + j_3 \beta + F_l \gamma_l + 2 \pi i \kappa (2 j_3+ R)\right], ~~~~ n \in \mathbb{Z},
\end{align}
where $w(u)$ denotes the weight of the given representation evaluated on the holonomy $u$.
The product of all eigenvalues \eqref{eq:Dphi} yields the determinat:
\begin{align}
    \det D_{\phi} & = \prod_{\rho \in \text{R}_{\phi}} \prod_{j = \frac{|\rho(m)|}{2}}^{\infty} \prod_{j_3 = -j}^{j}\prod_{n= -\infty}^{\infty} (j + r_{\phi} + \Delta_3)(j+1 - r_{\phi} - \Delta_3).  \label{eq:detDphi}
\end{align}
Let us separately analyze the factor $(\beta + \beta^{\prime}) \left(j + r_{\phi}+  \Delta_3\right)$ in \eqref{eq:detDphi} which is explicitly given as:
\begin{align}
 ( \beta+ \beta^{\prime}) \left(j + r_{\phi}+  \Delta_3\right)  = 2 \pi i n - i w(u)+ (j - j_3) \beta^{\prime} +  (j +r_{\phi} +j_3)  \beta + F_l \gamma_l +  2 \pi i \kappa (2 j_3+R). 
  \end{align}
  Let us define $z= (\beta+ \beta^{\prime}) \left(j + r_{\phi}+  \Delta_3\right)-2 \pi i n$ and perform the infinite product over all eigenvalues:
\begin{align}
\begin{split}
    \prod_{\text{all but }', n}\prod_{n= - \infty}^{\infty} (2 \pi i n +z)^{- (-1)^{F_j}} &=   \prod_{\text{all but }\, n} e^{-\frac{z}{2}(-1)^{F_j}} \exp \left[(-1)^{F_j}\sum_{m=1}^{\infty}\frac{1}{m}e^{- m z}\right] \\
    & =e^{-\sum_{\text{all but}\, n}\frac{z}{2}(-1)^{F_j}} \exp \left[\sum_{m=1}^{\infty}\frac{1}{m}\sum_{\text{all but}\, n}(-1)^{F_j}e^{- m z}\right].
    \end{split}
\end{align}
Here the notation ``all but $n$'' means that the product runs over all quantum numbers except $n$ that we have conveniently separated in order to massage the expressions.
Defining the single letter index as:
\begin{align}
\begin{split}
    f_z(u, \beta, \beta^{\prime}, \gamma_l) & = \sum_{\text{all but}\, n}(-1)^{F_j}e^{-  z} \\
    & =  \sum_{\text{all but}\, n}(-1)^{F_j} e^{i  w(u)} e^{- \beta^{\prime}(j - j_3)}e^{- \beta (j+ r_{\phi} + j_3)} e^{-  F_l \gamma_l}e^{-2 \pi i  \kappa (2 j_3 + R) }
    \end{split}
\end{align}
Making the indices over which we sum explicit, we have:
\begin{align}
    \begin{split}
          f_z(u, \beta, \beta^{\prime}, \gamma_l) & = \sum_{w \in \text{R}_{\phi}} e^{-i w(u)} e^{- \beta r_{\phi}-F_l \gamma_l }\sum_{j =\frac{|\rho(m)|}{2}}^{\infty} \sum_{j_3 = - j}^j e^{-(\beta^{\prime}+ \beta)(j)}e^{-(\beta- \beta^{\prime} )j_3}e^{- 2 \pi i \kappa (2 j_3+ R)},
    \end{split}
\end{align}

\begin{align}
    f_{\text{chiral}}(u, \beta, \gamma_l) =\sum_{\phi} \sum_{w \in \text{R}_{\phi}} \left[e^{i w(u)}e^{F_l \gamma_l} \frac{e^{- \beta(|w(\mathfrak{m})| + r_{\phi})}e^{- 2 \pi i \kappa r_{\phi}}}{1- e^{- 2\beta}}- e^{-i w(u)}e^{-F_l \gamma_l} \frac{e^{- \beta(|w(\mathfrak{m})| 2- r_{\phi})}e^{ 2 \pi i \kappa r_{\phi}}}{1- e^{- 2\beta}}\right].
\end{align}

In summary, the contribution to the 1-loop determinant can be written as 
\begin{align}
Z_{\text{1-loop}}^{\text{chiral}} = \prod_I \prod_{w \otimes \sigma \in \mathfrak{R}_I} \left(s^w \mathfrak{t}^\sigma q^{r_\sigma - 1}\right)^{-\frac{|w(\mathfrak{m}) + \sigma(\mathfrak{n})|}{2}} \frac{\left(s^{-w} \mathfrak{t}^{-\sigma} q^{2 - r_\sigma + |w(\mathfrak{m}) + \sigma(\mathfrak{n})|};q^2\right)}{\left(s^{w} \mathfrak{t}^{\sigma} q^{r_\sigma + |w(\mathfrak{m}) + \sigma(\mathfrak{n})|};q^2\right)} \label{eq:ZchiralShift},
\end{align}
where the first product runs over the chiral multiplets present, indexed by $I$; the second product runs over the gauge weights $w$ and the global weights $\sigma$ of the representation $\mathfrak{R}_I$ under the gauge and global symmetries, respectively.  Note that we have now  used somehow standard notation where $s_a=e^{iu_a}$ and $\mathfrak{t}$ is the global flavor fugacity. Other contributions can be obtained in a similar manner and we simply spell out the main results here.

The classical term $Z_{\text{class}}$, receives contributions from the Chern-Simons terms and any topological symmetry of the theory. A level $k$ canonical Chern-Simons term contributes
\begin{align}\label{Eq:CS-classical}
Z_{\text{class}}^{\text{CS}} = s^{k \mathfrak{m}},
\end{align}
a $U(1)$ topological symmetry with holonomy $\xi$ and flux $\mathfrak{t}$ will contribute \begin{align}
Z_{\text{class}}^{\text{top}} = s^\mathfrak{t} \xi^\mathfrak{m}.
\end{align}
The 1-loop determinant from the vector multiplet (gauge field) is 
\begin{align}
Z_{\text{1-loop}}^{\text{gauge}} = \prod_{\alpha \in \Delta} q^{-\frac{|\alpha(\mathfrak{m})|}{2}}\left(1 - s^\alpha q^{|\alpha(\mathfrak{m})|}\right) \label{eq:ZgaugeGuess},
\end{align}
where the product runs over the set $\Delta$ of non-zero roots $\alpha$ of the gauge group $G$. 



\section{Large-$N$, Cardy-like expansion of the ABJM SCI}\label{Sec:ABJM}
The ABJM theory is a Chern-Simons matter theory with gauge group  $U(N)_k \times U(N)_{-k}$, where the Chern-Simons levels of the gauge nodes are $k$ and $-k$. The theory has ${\cal N}=8$ supersymmetry for $k=1,2$ and ${\cal N}=6$ supersymmetry for generic $k$ \cite{Aharony:2008ug}.  In the 3d $\mathcal{N} = 2$ language of this paper, there are also 4 chiral multiplets, two of which transform in the bifundamental representation $(N,\bar{N})$ of the product gauge group, and two of which transform in the antibifundamental representation $(\bar{N}, N)$. Putting these data together with the expressions for the index components, the master index (\ref{Eq:3dSCI}) can be written as  \cite{Choi:2019zpz,Nian:2019pxj}:

\begin{align}\label{Eq:ABJM-Master}
    Z_{ABJM} = \frac{1}{(N!)^2}& \sum_{m_a, \tilde{m}_a \in \mathbb{Z}^N} \oint \left(\prod_{a = 1}^N \frac{ds_a}{2\pi i s_a} \frac{d\tilde{s}_a}{2 \pi i \tilde{s}_a} s_a^{k m_a} \tilde{s}_a^{-k \tilde{m}_a}\right) \nonumber \\
    &\prod_{a \not= b} q^{-\frac{1}{2}|m_{ab}|} \left(1 - s_a s_b^{-1} q^{|m_{ab}|}\right) \prod_{a \not= b} q^{-\frac{1}{2}|\tilde{m}_{ab}|}\left(1 - \tilde{s}_a \tilde{s}_b^{-1} q^{|\tilde{m}_{ab}|}\right) \nonumber \\
    &\prod_{I = 1,2} \prod_{a,b} q^{\frac{1}{4}|m_a - \tilde{m}_b|} \frac{(s_a^{-1} \tilde{s}_b t_I^{-1} q^{\frac{3}{2} + |m_a - \tilde{m}_b|};q^2)}{(s_a \tilde{s}_b^{-1} t_I q^{\frac{1}{2} + |m_a - \tilde{m}_b|};q^2)} \nonumber \\
    &\prod_{I = 3,4} \prod_{a,b} q^{\frac{1}{4}|-m_a + \tilde{m}_b|} \frac{(s_a \tilde{s}_b^{-1} t_I^{-1} q^{\frac{3}{2} + |-m_a + \tilde{m}_b|};q^2)}{(s_a^{-1} \tilde{s}_b t_I q^{\frac{1}{2} + |-m_a + \tilde{m}_b|};q^2)}.
\end{align}
The first line contains the classical contribution coming from the Chern-Simons terms (see equation (\ref{Eq:CS-classical})), the second line is the one-loop contribution of the vector multiplets (see equation (\ref{eq:ZgaugeGuess})) and the last two lines represent the contribution of the chiral multiplets  (see equation  (\ref{eq:ZchiralShift})). 
\subsection{Exact preliminaries}
Our ultimate goal is to better understand a systematic evaluation of the SCI given by the master equation (\ref{Eq:3dSCI}) in the particular case of the ABJM theory given in (\ref{Eq:ABJM-Master}). We will present explicit results after taking the large $N$ limit and further the Cardy-like expansion. However, to clarify the validity of the expansions we first perform a series of exact manipulations of this index.

Our strategy for evaluating the index will be to turn it into an expression amenable to a saddle point approximation. First, since this index is invariant under permutations of $m_a$ and $\tilde{m}_a$, if we fix both sequences to be non-decreasing in $a$, which we denote as $m_a, \tilde{m}_a \in \mathbb{Z}_d^N$, we can cancel the Weyl factor $(N!)^2$.  We choose a treatment that turns this discrete sum into a continuous integral by introducing a change of variable 
\begin{align}
m_a \rightarrow \frac{N^\alpha}{\beta} x_a, \label{eq:mToXRule}
\end{align}
for some yet undetermined $\alpha \in (0,1)$. The resulting $x_a \in \frac{\beta}{N^\alpha}\mathbb{Z}$ can now take on a continuum of values in either of the $N \rightarrow \infty$ or $\beta \rightarrow 0$ limits we consider. 

The choice of prefactor is motivated by the numerical observation that in the large $N$ limit, the values of the set $\{m_a\} \in \mathbb{Z}^N$ which extremize $Z_{SCI}$ (i.e. the saddle point) have a power law dependence $m_a \sim N^\alpha$. For ABJM, this is $\alpha = \frac{1}{2}$. The evidence for the value of $\alpha$ arises in direct analytical and numerical analyses performed, for example, in  \cite{Benini:2015eyy,Liu:2017vll} although one should note that originally it was  not directly carried for the fluxes, rather for the holonomies which were tacitly assumed to be complex. A more direct derivation of $\alpha=1/2$ for the superconformal index was presented in \cite{Choi:2019zpz,Nian:2019pxj}. Therefore, in the large $N$ limit, the corresponding saddle point values $x_a$ are of order $O(N^0)$, which will help us clarify which terms are leading in the large $N$ limit. 

Likewise, after we have taken the Large $N$ limit, and considered the Cardy limit, the leading scaling of the saddle point fluxes $m_a$ will be $\beta^{-1}$. So again the factor in (\ref{eq:mToXRule}) is chosen with that foresight so that the scaling of $x_a$ in the Cardy limit is $O(\beta^0)$. 

We also perform a few substitutions: \begin{align}
    s_a = e^{i y_a}, \quad q = e^{-\beta}, \quad t_I = e^{i \lambda_I}, \label{eq:fug}
\end{align} 
along with corresponding substitutions for $\tilde{m}_a$ and $\tilde{s}_a$. Note that we have notated the holonomies as $y_a$ instead of $u_a$ to match the usual presentation of the topologically twisted index. The original fugacities $t_I$ must satisfy $|t_I| = 1$ and $t_1 t_2 t_3 t_4 = 1$, so we have real-valued $\lambda_I = \log t_I \in [0,2 \pi)$ and \begin{align}
    \sum_I \lambda_I \in \{0, 2\pi, 4 \pi, 6 \pi\}.
\end{align}
We will focus on the $\sum_I \lambda_I = 2 \pi$ case motivated by the fact that for the topologically twisted index it is related to the $\sum_I \lambda_I = 6 \pi$ case, and the $\sum_I \lambda_I = 0, 4 \pi$ cases yield trivial/no solutions at large $N$.

With these changes of variables, we can write the index exactly in $N$ and $\beta$ as 
\begin{align}
    Z_{ABJM} = \sum_{x_a, \tilde{x}_a \in \frac{\beta}{N^\alpha}\mathbb{Z}_d^N}\left(\prod_{a=1}^N \int_0^{2 \pi} \frac{d y_a}{2 \pi}\frac{d \tilde{y}_a}{2 \pi}\right) \exp \left(-S[y_a,\tilde{y}_a, x_a, \tilde{x}_a]\right).
\end{align}

Since we consider $\beta$ with $\text{Re}[\beta] > 0$, we can write the q-Pochhammer symbols out in a convergent expansion, leading to an action of the form: 
\begin{align}
S[y_a,\tilde{y}_a,x_a,\tilde{x}_a] &= N^\alpha \sum_{a,b=1}^N \left[|x_a - \tilde{x}_b| - \frac{1}{2}|x_a - x_b| - \frac{1}{2}|\tilde{x}_a - \tilde{x}_b|\right] \nonumber \\
&+ \sum_{a = 1}^N ik \frac{N^\alpha}{\beta} \left( x_a y_a - \tilde{x}_a \tilde{y}_a \right) \nonumber\\
&+ \sum_{a \not= b} \left[ \text{Li}_1 \left(e^{iy_{ab}-N^\alpha|x_{ab}|}\right) + \text{Li}_1 \left(e^{i\tilde{y}_{ab}-N^\alpha|\tilde{x}_{ab}|}\right)\right] \nonumber\\
&+ \sum_{A = 1,2} \sum_{n=0}^\infty \sum_{a,b = 1}^N \Bigg[\text{Li}_1\left(e^{i(\tilde{y}_b - y_a - \lambda_A) - \frac{3}{2} \beta-2\beta n-N^\alpha |x_a - \tilde{x}_b|}\right) \nonumber\\
&\quad\quad\quad\quad\quad\quad\quad\quad\quad- \text{Li}_1 \left(e^{-i(\tilde{y}_b - y_a - \lambda_A) - \frac{1}{2} \beta-2\beta n-N^\alpha |x_a - \tilde{x}_b|}\right) \Bigg] \nonumber\\
&+ \sum_{B = 3,4} \sum_{n=0}^\infty \sum_{a,b = 1}^N \Bigg[\text{Li}_1 \left(e^{i(y_a - \tilde{y}_b - \lambda_B) - \frac{3}{2} \beta-2\beta n-N^\alpha |-x_a + \tilde{x}_b|} \right) \nonumber \\
&\quad\quad\quad\quad\quad\quad\quad\quad\quad- \text{Li}_1 \left(e^{-i(y_a -\tilde{y}_b - \lambda_B) - \frac{1}{2} \beta-2\beta n-N^\alpha |-x_a + \tilde{x}_b|}\right) \Bigg]. \label{eq:SABJMExact}
\end{align}
This is most grotesque form of the index we will encounter, but it is still exact in $N$ and $\beta$.

\subsection{Large $N$ Limit}
In the large $N$ limit, the first line of the expression (\ref{eq:SABJMExact}) for the full action could in principle grow as $N^{2 + \alpha}$ for arbitrary choice of $x_a$ and $\tilde{x}_a$: \begin{align}
   O(N^{2 + \alpha}): \quad N^\alpha \sum_{a,b=1}^N \left[|x_a - \tilde{x}_b| - \frac{1}{2}|x_a - x_b| - \frac{1}{2}|\tilde{x}_a - \tilde{x}_b|\right].
\end{align} However, this term is minimized, and vanishes for only $x_a = \tilde{x}_a$, and so this $O(N^{2 + \alpha})$ term simply exactly enforces this at lower orders, a similar situation takes place in the topologically twisted index discussed analytically in  \cite{Benini:2015eyy} with further numerical details in \cite{Liu:2017vll}.

With this in mind, we can write the now effective large $N$ action as \begin{align}
\begin{split}
S_{eff}[x_a,y_a,\tilde{y}_a] &=  -\sum_{a = 1}^N ik \frac{N^\alpha}{\beta} x_a \left(\tilde{y}_a - y_a \right)\\
&+ \sum_{a \not= b} \left[ \text{Li}_1 \left(e^{iy_{ab}-N^\alpha|x_{ab}|}\right) + \text{Li}_1 \left(e^{i\tilde{y}_{ab}-N^\alpha|x_{ab}|}\right)\right]\\
&+ \sum_{A = 1,2} \sum_{n=0}^\infty \sum_{a,b = 1}^N \Bigg[\text{Li}_1\left(e^{i(\tilde{y}_b - y_a - \lambda_A) - \frac{3}{2} \beta-2\beta n-N^\alpha |x_{ab}|}\right)\\
&\quad\quad\quad\quad\quad\quad\quad\quad\quad- \text{Li}_1 \left(e^{-i(\tilde{y}_b - y_a - \lambda_A) - \frac{1}{2} \beta-2\beta n-N^\alpha |x_{ab}|}\right) \Bigg]\\
&+ \sum_{B = 3,4} \sum_{n=0}^\infty \sum_{a,b = 1}^N \Bigg[\text{Li}_1 \left(e^{i(y_a - \tilde{y}_b - \lambda_B) - \frac{3}{2} \beta-2\beta n-N^\alpha |x_{ab}|} \right) \\
&\quad\quad\quad\quad\quad\quad\quad\quad\quad- \text{Li}_1 \left(e^{-i(y_a -\tilde{y}_b - \lambda_B) - \frac{1}{2} \beta-2\beta n-N^\alpha |x_{ab}|}\right) \Bigg].
\end{split}\label{eq:SABJMLargeN1}
\end{align}
To specify a candidate choice of $\{x_a\} \in \frac{\beta}{N^\alpha} \mathbb{Z}^N$, it is convenient to defined a normalized density  $\rho(x) = \frac{1}{N}\sum_a \delta(x - x_a)$ in terms of which  we can rewrite sums over the indices as $\sum_a (...) \mapsto N \int dx \rho(x) (...)$. In the $N \rightarrow \infty$ limit, the possible choices of $\{ x_a \}$ are then parameterized by choices of functions of $\rho(x)$ subject to $\rho(x) \geq 0$ and $\int dx \rho(x) = 1$. An example of values of $x_a$ that correspond to a given distribution $\rho(x)$ in the $N \rightarrow \infty$ limit is given in (\ref{eq:xa-choice}).  

After this continuum representation of the monopole configurations, one still needs to evaluate the integral over holonomies, $s_a = e^{i y_a}$ in the master expression (\ref{Eq:ABJM-Master}). In the large-$N$ limit the discrete variables $y_a \in \mathbb{R}$ may be similarly determined by a function $y(x) \in \mathbb{R}$ that interpolates between these holonomies: $y_a = y(x_a)$. In fact, we will find that the integrand will depend only on the differences $\tilde{y}_a - y_a$, represented by a function $\delta y(x) = \tilde{y}(x) - y(x)$. The problem is then
\begin{align}
    Z_{ABJM} = \int \mathcal{D} \rho \mathcal{D} \delta y \exp \left(-S_{eff}[\rho, \delta y] \right), \label{eq:Z-S-eff}
\end{align}
where we can write $S_{eff}$ in terms of a local Lagrangian, grouped according to growth in $N$: \begin{align}
    S_{eff}[\rho, \delta y] = \int dx \left[N^{3/2}\mathcal{L}_{\text{ABJM}}^{(3/2)}(x,\rho(x),\delta y(x)) +  \mathcal{L}_{\text{ABJM}}^{(sub)}(x,\rho(x),\delta y(x))\right] + o(N^{3/2}). \label{eq:S-eff-form}
\end{align}

To evaluate the index (\ref{eq:Z-S-eff}), we can then perform a saddle point analysis in $\rho(x)$ and $\delta y(x)$ where we first find the saddle point values $\rho^*(x)$ and $\delta y^*(x)$ and then evaluate the action at that point. To find the saddle point, we must still ensure that the resulting distribution $\rho^*(x)$ is normalized. This can be done by introducing a Lagrange multiplier term into the action: \begin{align}
    S_\mu[\rho,\delta y] = N^{3/2} \frac{i k}{\beta}\,\, \mu\,\, \left(\int dx \rho(x) -1\right). \label{eq:mu-Term}
\end{align}
During the saddle point analysis, we can then solve for $\mu$ so that the resulting distribution $\rho^*(x)$ is normalized. The choice of prefactor of $\mu$ is for later convenience. 

Although we are only concerned with the leading in $N$ result for the index, we must keep track of the subleading in $N$ term $\mathcal{L}_{\text{ABJM}}^{(sub)}$ since it has a significant impact on the ``tails" of the saddle point values $\rho^*(x)$ and $\delta y^*(x)$ in a way analogous to the large-$N$ treatment of the topologically twisted index of ABJM  in \cite{Benini:2015eyy}. This occurs due to singularities in $\delta y(x)$ present in the local Lagrangian, which enable the $\mathcal{L}_{\text{ABJM}}^{(sub)}$ term to compete with the leading $O(N^{3/2})$ behavior if $\delta y(x)$ approaches the singularity as $N \rightarrow \infty$ (as they will). Because of this effect, we also only need to ensure that $\mathcal{L}_{\text{ABJM}}^{(sub)}$ reflects any singularities in $\rho(x)$ and $\delta y(x)$, smooth subleading behavior is irrelevant for our analysis. The influence of these singularities is clarified in Section \ref{Sec:SPApproximation}.

The first line of (\ref{eq:SABJMLargeN1}), the classical contribution, can be treated directly as \begin{align}
    -\sum_{a=1}^N i k \frac{N^\alpha}{\beta} x_a (\tilde{y}_a - y_a) = - N^{3/2} \frac{i k}{\beta}\int dx \rho(x) x \delta y(x).
\end{align}

To make progress on the other terms, which are double sums, we use large $N$ ``localization" arguments detailed in Appendix \ref{AppNLocal}. These tell us that if we assume that the choices of $x_a$ converge in the large $N$ limit to some distributions $\rho(x)$, we can perform the following general substitution at large $N$, (\ref{eq:Li-1-DS}): \begin{align}
  &\sum_{a=1}^N \sum_{b=1}^N \text{Li}_1 \left(e^{i(h_1(x_a) + h_2(x_b)) - N^\alpha |x_{ab}|}\right) \\
      &= 2 N^{2 - \alpha} \int dx \rho(x)^2 \text{Li}_2\left(e^{i(h_1(x)+h_2(x))}\right) - \frac{N^{\alpha}}{6} \int dx \text{Li}_0\left(e^{i(h_1(x)+h_2(x))}\right) \\
    &- N^{2 - 2\alpha}\left(\rho(x_1)^2\text{Li}_3 \left(e^{i(h_1(x_1)+h_2(x_1))}\right)+\rho(x_N)^2\text{Li}_3 \left(e^{i(h_1(x_N)+h_2(x_N))}\right)\right).
\end{align}
For our purposes ($\alpha = \frac{1}{2}$), this expression is valid up to $O(N)$. 

For example, if we choose $h_1(x_a) = -y_a$ and $h_2(x_b) = \tilde{y}_b - \lambda_A + i\frac{3}{2} \beta + i 2 \beta n$, the large $N$ expansion of a portion of the chiral multiplet contribution on the third line of (\ref{eq:SABJMLargeN1}) is \begin{align}
    &\sum_{a,b = 1}^N \text{Li}_1\left(e^{i(\tilde{y}_b - y_a - \lambda_A) - \frac{3}{2} \beta-2\beta n-N^\alpha |x_{ab}|}\right) \\
    &= 2 N^{\frac{3}{2}} \int dx \rho(x)^2 \text{Li}_2\left(e^{i(\delta y(x) - \lambda_A)- \frac{3}{2} \beta-2\beta n}\right) - \frac{N^{\frac{1}{2}}}{6} \int dx \text{Li}_0\left(e^{i(\delta y(x) - \lambda_A)- \frac{3}{2} \beta-2\beta n}\right) \\
    &- N\left(\rho(x_1)^2\text{Li}_3 \left(e^{i(\delta y(x_1) - \lambda_A) - \frac{3}{2} \beta-2\beta n}\right)+\rho(x_N)^2\text{Li}_3 \left(e^{i(\delta y(x_N) - \lambda_A) - \frac{3}{2} \beta-2\beta n}\right)\right). \label{eq:example-bottom-line}
\end{align}
The first part of this term contributes to the leading Lagrangian $\mathcal{L}_{\text{ABJM}}^{(3/2)}$. The second part contributes to $\mathcal{L}_{\text{ABJM}}^{(sub)}$, since it generates singularities in $\delta y(x)$ given that $\text{Li}_0(e^{iz})\sim\frac{i}{z}$. The bottom line (\ref{eq:example-bottom-line}) does not contribute to either since it is subleading in $N$ yet does not generate singularities. In fact, we will find that the saddle point distribution for $\rho(x)$ vanishes on the endpoints $x_1$ and $x_N$, so this term would vanish as well regardless. 

The second line of (\ref{eq:SABJMLargeN1}), the gauge node contribution, has some further subtleties to consider so that we cannot simply apply (\ref{eq:Li-1-DS}). Further discussion of these details may be found in the Appendix surrounding (\ref{eq:Li-1-DSNE}). The resulting contribution to our Lagrangian is: 
\begin{align}
    \mathcal{L}_{\text{gauge}}^{(3/2)}(x, \rho(x), \delta y(x)) &= N^{3/2} \rho(x)^2 \frac{\pi^2}{3}.
\end{align}

Combining these results, we find the full form of the large $N$ effective Lagrangian, with terms grouped according to their scaling in $N$ as in (\ref{eq:S-eff-form}): \begin{align}
    \mathcal{L}_{\text{ABJM}}^{(3/2)}(x,\rho,y) &= -\frac{i k}{\beta} \rho \left(x y - \mu \right) \label{eq:L-ABJM-3-2}\\
    &+ 2\rho^2 \Big(\frac{\pi^2}{3} + \sum_{n=0}^\infty \Bigg[\text{Li}_2\left(e^{i(y - \lambda_{1,2}) - \frac{3}{2} \beta - 2 \beta n}\right) - \text{Li}_2\left(e^{-i(y - \lambda_{1,2}) - \frac{1}{2} \beta - 2 \beta n}\right) \nonumber\\
    &\quad\quad\quad+\text{Li}_2\left(e^{i(-y - \lambda_{3,4}) - \frac{3}{2} \beta - 2 \beta n}\right) - \text{Li}_2\left(e^{-i(-y - \lambda_{3,4}) - \frac{1}{2} \beta - 2 \beta n}\right)\Bigg]\Big) \nonumber\\
    \mathcal{L}_{\text{ABJM}}^{(sub)}(x,\rho,y) &= -\frac{N^{\frac{1}{2}}}{6}\sum_{n=0}^\infty \Bigg[\text{Li}_0\left(e^{i(y - \lambda_{1,2}) - \frac{3}{2} \beta - 2 \beta n}\right) - \text{Li}_0\left(e^{-i(y - \lambda_{1,2}) - \frac{1}{2} \beta - 2 \beta n}\right) \label{eq:L-ABJM-1} \\
    &\quad\quad\quad+\text{Li}_0\left(e^{i(-y - \lambda_{3,4}) - \frac{3}{2} \beta - 2 \beta n}\right) - \text{Li}_0\left(e^{-i(-y - \lambda_{3,4}) - \frac{1}{2} \beta - 2 \beta n}\right)\Bigg] 
\end{align}
where we have rewritten $y = \delta y (x)$, $\rho = \rho(x)$ for brevity, as we will in subsequent sections. The Lagrange multiplier term $\mu$ from (\ref{eq:mu-Term}) has also been absorbed into the definition of $ \mathcal{L}_{\text{ABJM}}^{(3/2)}(x,\rho,y)$.

Critically, we emphasize that this is only the large $N$ Lagrangian, and is still valid at finite $\beta$. This Lagrangian therefore captures nonperturbative effects from the perspective of the Cardy-like limit $\beta \rightarrow 0$ and could be analyzed at finite $\beta$. However, the saddle point analysis that we use is only analytic and tractable in the Cardy-like limit, which we explore in the following section. 

From the subleading term in the Lagrangian (\ref{eq:L-ABJM-1}), we can identify singularities in $y$ present at \begin{align}
    y = \lambda_{1,2} + \frac{i \beta}{2} + 2\beta i n, \quad  y = - \lambda_{3,4} - \frac{i \beta}{2} +  2 \beta i n,  \quad n \in \mathbb{Z}.
\end{align} 

Ultimately, these divergences will mean that it is possible for the leading saddle point solution to get ``stuck" at any of these subleading singularities and form a tail, analogous to the same effect in the large $N$ solution of the topologically twisted index. 

\subsection{Cardy-like Limit}
As mentioned in the last section, we will focus our attention on the Cardy-like limit of the leading in $N$ term, $\mathcal{L}_{\text{ABJM}}^{(3/2)}(x,\rho,y)$. For this, we use the identity (\ref{eq:Li2-Identity}) demonstrated in the Appendix: \begin{align}
    \sum_{n=0}^\infty \text{Li}_2 \left(z q^{a + 2n}\right) = \frac{1}{2\beta}\sum_{r=0}^\infty B_r\left(1 - \frac{a}{2}\right) \frac{(2 \beta)^r}{r!} \text{Li}_{3-r}\left(z\right),
\end{align}
to write the Cardy-like limit of the leading large $N$ term as \begin{align}
&\mathcal{L}_{\text{ABJM}}^{(3/2)}(x,\rho,y) \nonumber \\
& = -\frac{ik}{\beta} \rho(x y - \mu) \\
& \quad + 2 \rho^2 \Big(\frac{\pi^2}{3} + \frac{1}{2 \beta}\sum_{r=0}^\infty \frac{( 2\beta)^r}{r!}\Big[ B_r\left(\frac{1}{4}\right)\text{Li}_{3 - r}\left(e^{i(y - \lambda_{1,2})}\right)+B_r\left(\frac{3}{4}\right)\text{Li}_{3 - r}\left(e^{-i(y - \lambda_{1,2})}\right) \nonumber\\
& \quad \quad \quad \quad\quad\quad\quad+B_r\left(\frac{1}{4}\right)\text{Li}_{3 - r}\left(e^{i(-y - \lambda_{3,4})}\right) + B_r\left(\frac{3}{4}\right)\text{Li}_{3 - r}\left(e^{-i(-y - \lambda_{3,4})}\right)\Big]\Big)\nonumber\\
& = -\frac{ik}{\beta} \rho(x y - \mu) \\
& \quad + 2 \rho^2 \Big(\frac{\pi^2}{3} + \frac{1}{2 \beta}\sum_{r=0}^\infty B_r\left(\frac{1}{4}\right) \frac{( 2\beta)^r}{r!}\Big[ \text{Li}_{3 - r}\left(e^{i(y - \lambda_{1,2})}\right)+(-1)^{3-r}\text{Li}_{3 - r}\left(e^{-i(y - \lambda_{1,2})}\right) \nonumber\\
& \quad \quad \quad \quad\quad\quad\quad+\text{Li}_{3 - r}\left(e^{i(-y - \lambda_{3,4})}\right) + (-1)^{3-r}\text{Li}_{3 - r}\left(e^{-i(-y - \lambda_{3,4})}\right)\Big]\Big)\nonumber
\end{align}
To get the last line, we have first used that $B_r(a) = (-1)^r B_r(1-a)$. This then introduces the appropriate signs to use the polylogarithm reflection identity: 
\begin{align}
   \text{Li}_n(e^{ia}) + (-1)^n \text{Li}_n(e^{-ia}) = -\frac{(2 \pi i)^n}{n!} B_n\left(\frac{a}{2\pi} - \left\lfloor \frac{\text{Re}(a)}{2\pi}\right\rfloor \right) \equiv P_{(n)}(a). \label{eq:Pna}
\end{align}
In practice, we will consider arguments $a$ where $\text{Re}(a) \in [-2 \pi, 0)$, so that the relevant $P_{(n)}(a)$ are \begin{align}
    &P_{(0)}(a) = -1, \qquad P_{(1)}(a) = -i(a + \pi) \\
    &P_{(2)}(a) = \frac{a^2}{2} + a \pi + \frac{\pi^2}{3}, \qquad P_{(3)}(a) = i \left(\frac{a^3}{6} + \frac{a^2 \pi}{2} + \frac{a \pi^2}{3}\right)
\end{align}
Crucially, these $P_{(n)}(a)$ are polynomials, and so for $n < 0$, $P_{(n)}(a) = 0$ and the Cardy-like expansion truncates at $O(\beta^2)$: \begin{align}
    &\mathcal{L}_{\text{ABJM}}^{(3/2)}(x,\rho,y) = -\frac{ik}{\beta} \rho(x y - \mu) \nonumber \\
& \quad + 2 \rho^2 \Big(\frac{\pi^2}{3} + \frac{1}{2 \beta}\sum_{r=0}^3 B_r\left(\frac{1}{4}\right) \frac{( 2\beta)^r}{r!}\Big[ P_{(3-r)}(y - \lambda_{1,2}) + P_{(3-r)}(-y - \lambda_{3,4})\Big] \Big).
\end{align}

Although we have observed this truncation in $\beta$, we emphasize that the even the leading Lagrangian (\ref{eq:L-ABJM-3-2}) contains nonperturbative aspects which this treatment neglects. 

Issues arise if we attempt to extend our Cardy-like expansion to the terms which are subleading in $N$. If we apply the expansion of the q-Pochammer symbol (\ref{Eq:Pochhamer-IdentityMain}), appropriately differentiated or integrated, we can take the Cardy-like limits of the sums of polylogarithms which appear. However, if the index of the polylogarithm is odd, as it can be in (\ref{eq:Li-1-DS}) and other subleading terms, we end up with the wrong sign between the polylogarithms to apply the reflection identity (\ref{eq:Pna}). There is thus no truncation in $\beta$ and the expansion of the Lagrangian does not converge, just as the expansion of the q-Pochammer symbol does not converge in Fig.~\ref{fig:Pochhammer-pert-approx}. For this reason, in our analysis it has been important to observe the singularities of the subleading term in (\ref{eq:L-ABJM-1}) before taking the Cardy-like limit. Then we may move forward with only the well-behaved $O(N^{3/2})$ term. 

\subsection{Saddle Point Approximation}\label{Sec:SPApproximation}

From the previous two sections, we have seen that the Cardy-like expansion of the leading in $N$ term of our effective Lagrangian truncates as \begin{align}
 &\mathcal{L}_{\text{ABJM}}^{(3/2)}(x,\rho,y) = -\frac{ik}{\beta} \rho(x y - \mu) \nonumber \\
& \quad + 2 \rho^2 \Big(\frac{\pi^2}{3} + \frac{1}{2 \beta}\sum_{r=0}^3 B_r\left(\frac{1}{4}\right) \frac{( 2\beta)^r}{r!}\Big[ P_{(3-r)}(y - \lambda_{1,2}) + P_{(3-r)}(-y - \lambda_{3,4})\Big] \Big).
\end{align}

From the large-$N$ analysis alone, we note that the subleading in $N$ term also contains singularities in $y$. For the case $\sum_a \lambda_a = 2 \pi$ that we consider, the relevant singularities are those at \begin{align}
    \delta y(x) = \lambda_{1,2} + \frac{i \beta}{2}, \quad \delta y(x) = - \lambda_{3,4} - \frac{i \beta}{2}.
\end{align}
Motivated by this, we will introduce shifts \begin{align}
    \Delta_a = \lambda_a + \frac{i \beta}{2}, \quad \sum_a \lambda_a = 2 \pi + 2 i \beta. \label{eq:Delta-lambda-shift}
\end{align} so that these singularities now arise at \begin{align}
    \delta y(x) = \Delta_{1,2}, \quad \delta y(x) =-\Delta_{3,4}
\end{align}

We will also take advantage of our definitions of the $\lambda_i$ up to multiples of $2\pi$ to assume that \begin{align}
    -\lambda_4 \leq -\lambda_3 \leq \text{Re}[\delta y(x)] \leq \lambda_1 \leq \lambda_2.
\end{align}
The real part of the saddle point solution for $\delta y(x)$ will then be restricted to the range $[-\lambda_3, \lambda_1]$ due to the singularities. 

The shifts (\ref{eq:Delta-lambda-shift}) also resonate with a recent treatment of the refined topologically twisted index where, to leading order,  shifts  by $\beta$ in the chemical potentials allow for a succinct writing of the answer \cite{Hosseini:2022vho} in agreement with the principle of holomorphic block factorization \cite{Beem:2012mb} for three-dimensional partition functions. In terms of these fugacities, the effective Lagrangian is \begin{align}
\mathcal{L}_{\text{ABJM}}^{(3/2)}(x,\rho,y) & = \frac{i}{\beta} \left[k \rho(\mu - x y) + \rho^2 \left(\pi y^2 -y(\Delta_1 \Delta_2 - \Delta_3 \Delta_4) - \frac{1}{2} \sum_{A < B < C} \Delta_A \Delta_B \Delta_C\right)\right] \nonumber\\
&- y^2 \rho^2 - \frac{2}{3} i \pi \beta \rho^2.
\end{align}
From saddle point analysis of just this leading term, we can find the ``inner solution" for $\rho(x)$ and $\delta y(x)$, now written out as functions of $x$: \begin{align}
\rho^{(i)}(x) &= \frac{k(2(\pi + i \beta)\mu - x(\Delta_1 \Delta_2 -\Delta_3 \Delta_4))}{\frac{8}{3} \pi (\pi + i \beta)\beta^2 + (\Delta_1 + \Delta_3)(\Delta_2 + \Delta_3)(\Delta_1 + \Delta_4)(\Delta_2 + \Delta_4)},\\
\delta y^{(i)}(x) &= \frac{\frac{4}{3} \pi x \beta^2 + \mu(\Delta_1 \Delta_2 - \Delta_3 \Delta_4) + x \sum_{A < B< C} \Delta_A \Delta_B \Delta_C}{2(\pi + i \beta)\mu - x(\Delta_1 \Delta_2 -\Delta_3 \Delta_4)}.
\end{align}
If we assume that on the left and right tails, the solution approaches $y = -\Delta_3$ and $y = \Delta_1$ respectively, as we have reason to believe from the finite $\beta$ expression, we then find the left tail solution: \begin{align}
\delta y^{(l)}(x) = -\Delta_3 + O(e^{-N}), \qquad \rho^{(l)}(x) = \frac{k(\mu + x \Delta_3)}{\frac{4}{3} \pi \beta^2 + (\Delta_1 + \Delta_3)(\Delta_2 + \Delta_3)(\Delta_4 - \Delta_3)},
\end{align}
and the right tail solution: \begin{align}
\delta y^{(r)}(x) = \Delta_1 + O(e^{-N}), \qquad \rho^{(r)}(x) = \frac{k(\mu - x \Delta_1)}{\frac{4}{3} \pi \beta^2 + (\Delta_1 + \Delta_4)(\Delta_1 + \Delta_3)(\Delta_2 - \Delta_1)}.
\end{align}

With these, we can then find (complex) values of $x$ at which these expressions agree: \begin{align}
\rho^{(l)}(x_\ll) = 0 \quad &\rightarrow \quad x_\ll = -\frac{\mu}{\Delta_3},
\end{align}
\begin{align}
\rho^{(l)}(x_<) = \rho^{(i)}(x_<) \quad &\rightarrow \quad x_< = -\frac{\mu}{\Delta_4 + \frac{4 \pi \beta^2}{3(\Delta_1 + \Delta_3)(\Delta_2 + \Delta_3)}},
\end{align}
\begin{align}
\rho^{(r)}(x_>) = \rho^{(i)}(x_>) \quad &\rightarrow \quad x_> = \frac{\mu}{\Delta_2 + \frac{4 \pi \beta^2}{3(\Delta_1 + \Delta_3)(\Delta_1 + \Delta_4)}},
\end{align}
\begin{align}
\rho^{(r)}(x_\gg) = 0 \quad \rightarrow \quad x_\gg = \frac{\mu}{\Delta_1}.
\end{align}

\begin{figure}[t]
\centering
\includegraphics[width=0.7\textwidth]{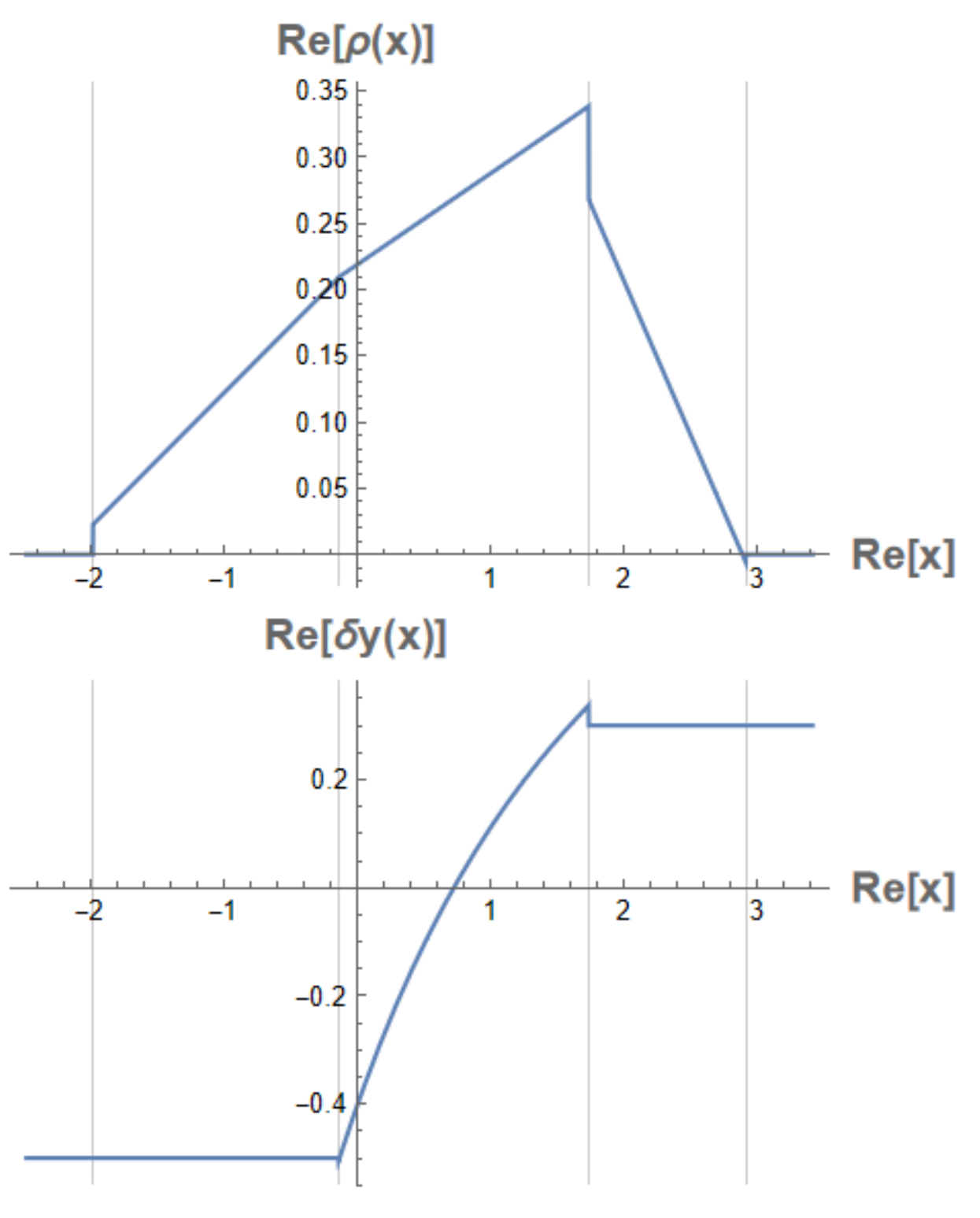}
\caption{Example of the saddle point solution in the large $N$, Cardy-like expansion. Values are $\lambda_1 = 0.3, \lambda_2 = 0.4, \lambda_3 = 0.5, \lambda_4 = 2 \pi - 1.2$, and $\beta = 0.5$. The regions (left-tail, center, right-tail) of $x$ are denoted by vertical lines. Only the real parts of $x$ and the solutions are plotted. The apparent discontinuities are only a result of this real projection.}\label{fig:saddle-point-example}
\end{figure}

Figure \ref{fig:saddle-point-example} shows example saddle point distributions of $\rho(x)$ and $\delta y(x)$. Generally, the saddle point distribution $\rho(x)$ has a piecewise linear form, and $\delta y(x)$ has a left-tail region at the singularity $-\Delta_3$, a non-linear inner section, and a right-tail region at the singularity $\Delta_1$. At leading order $O(\beta^{-1})$, all of the endpoints in $x$ that delineate the regions and the saddle point solutions are real. When the full expansion is considered, this is no longer the case. Since Figure \ref{fig:saddle-point-example} plots the real components of $x$ and of the solutions, the solutions appear discontinuous, although they do (by definition) in fact align at the appropriate complex values of $x$ written above. 

With these data we evaluate the index (\ref{eq:Z-S-eff}) as: 
\begin{align}
\log Z_{ABJM} &= -N^{3/2}\int_{-\infty}^\infty dx \int d \rho d y \mathcal{L}_{\text{ABJM}}^{(3/2)}(x,\rho,y) \\
&\approx -N^{3/2} \int_{-\infty}^\infty dx \mathcal{L}_{\text{ABJM}}^{(3/2)}(x,\rho^*(x),\delta y^*(x)).
\end{align}
The integral over $x$ is now deformed as a piecewise-linear contour between the complex endpoints $x_\ll, x_<, x_>, x_\gg$ identified earlier. Plugging in the various saddle point solutions yields a quite simple result: \begin{align}
    \log Z_{ABJM} &= -N^{3/2} \frac{i k}{3 \beta} \mu.
\end{align}
To evaluate the Lagrange multiplier $\mu$ we impose that $\int dx \rho^*(x) = 1$, again along the piecewise-linear complex contour. The resulting expression for $\mu$ gives the final result 
\begin{empheq}[box=\fbox]{equation}\label{eq:Z-ABJM-Final}
    \log Z_{ABJM} = -N^{3/2} \frac{i k\sqrt{\frac{2 }{3}}  \sqrt{\frac{ \Delta _1 \Delta _3 \left(4 \pi  \beta ^2+3 \left(\Delta _1+\Delta _3\right) \left(\Delta _2+\Delta _3\right) \Delta _4\right) \left(4 \pi  \beta ^2+3 \Delta _2 \left(\Delta _1+\Delta _3\right) \left(\Delta _1+\Delta _4\right)\right)}{\left(\Delta _1+\Delta _3\right)  \left(4 \pi  \beta ^2+3 \left(\Delta _1+\Delta _3\right) \left(\Delta _2+\Delta _3\right) \left(\Delta _1+\Delta _4\right)\right)}}}{3 \beta}.
\end{empheq}   
This is one of our main results in this paper, an evaluation of the superconformal index for the ABJM theory in the large-$N$ limit that is exact $\beta$ up to non-perturbative corrections. The precise argument for the nonperturbative corrections in $\beta$ was observed from the form of the large $N$ Lagrangian at finite $\beta$, (\ref{eq:L-ABJM-3-2}).

We note that this expression for the index does not truncate in the small $\beta$ expansion. Indeed, it takes the following form
 \begin{align}
       \log Z_{ABJM} &= -N^{3/2} \frac{i \sqrt{2 k \Delta_1 \Delta_2 \Delta_3 \Delta_4}}{3 \beta} \\
    &- N^{3/2} i \beta \frac{2 \pi \sqrt{2 k} \Delta_1 \Delta_3(\Delta_1 \Delta_2 + (\Delta_2+\Delta_3)\Delta_4)}{9(\Delta_1 + \Delta_3)(\Delta_2 + \Delta_3)(\Delta_1 + \Delta_4)\sqrt{\Delta_1 \Delta_2 \Delta_3 \Delta_4}} + O(\beta^2).  
\end{align}
It would be interesting to understand the above structure, particularly the higher order terms,  from the EFT point of view.

\section{The Cardy expansion truncates in the large $N$ limit}\label{Sec:LargeN}

In this section we demonstrate that by taking the large $N$ limit of the SCI first, we have a Cardy expansion that truncates, that is, only contains terms proportional to $1/\beta, \beta^{i=0,1,2}$ for a generic 3d ${\cal N}=2$ theory with gauge group of the form product of $U(N)$. In appendix \ref{qPochDev} we show that if we instead chose to take the Cardy expansion first, at finite $N$, the result is an uncontrollable expansion.

Using essentially the same preliminaries introduced for ABJM in the previous section, we now describe how the large-$N$ limit modifies the various building blocks of the SCI. The classical contribution is given by \begin{align}
   \log Z_{\text{class}}^{CS} &= k \mathfrak{m} \log(s) .
\end{align}
Explicitly, for $G = U(N)$, this is 
\begin{align}\label{Eq:Zgauge-a}
    \log Z_{\text{class}}^{U(N)} &= ik \frac{N^\alpha}{\beta}\sum_{a=1}^N x_a y_a = ik \frac{N^{1+\alpha}}{\beta}\int dx \rho(x) x y(x).
\end{align}

The gauge multiplet contribution is
\begin{align}
     \log Z_{\text{1-loop}}^{\text{gauge}} &= \sum_{\alpha \in \Delta} \left( \frac{N^\alpha}{2} |\alpha(x)| - \text{Li}_1(s^\alpha e^{-N^\alpha |\alpha(x)|})\right).
\end{align}
Or for the gauge group $G = U(N)$, then this is \begin{align}
    \log Z_{\text{1-loop}}^{U(N)} &= N^\alpha \left[\sum_{a < b} (x_b - x_a)\right] - N^2 \int dx dx' \rho(x)\rho(x') \text{Li}_1(s^\alpha e^{-N^\alpha |x - x'|}) \\
    &= N^\alpha \left[\sum_{a < b} (x_b - x_a)\right] - N^{2 - \alpha} \frac{\pi^2}{3} \int \rho(x)^2 dx + o(N^{3/2}).
\end{align}
The first line is just a formal manipulation into an integral, the second line is where the large $N$ limit first comes into play to localize the double integral. The details of this process can be found in appendix \ref{AppNLocal}.  

The trickiest term to deal with is the chiral multiplet.  For the contribution of a chiral multiplet, the Cardy-like expansion under similar assumptions,  is: 
\begin{align}\label{Eq:Zgauge-b}
    \log Z_{\text{1-loop},I}^{\text{chiral}} &= \log \left[\prod_{w \in \mathfrak{R}_I} \left(s^w \mathfrak{t}_I q^{r_I - 1}\right)^{-\frac{|w(\mathfrak{m})|}{2}} \frac{\left(s^{-w} \mathfrak{t}_I^{-1} q^{2 - r_I + |w(\mathfrak{m})|};q^2\right)}{\left(s^{w} \mathfrak{t}_I q^{r_I + |w(\mathfrak{m})|};q^2\right)}\right]\\
    &=  -\frac{N^\alpha}{2\beta}(i \lambda_I + \beta(1 - r_I))\sum_{w \in \mathfrak{R}_I}|w(x)|  \\
    -\frac{1}{2\beta}\sum_{r=0} \frac{(2 \beta)^r}{r!} &\sum_{w \in \mathfrak{R}_I}B_r\left(\frac{r_I}{2}\right)\Big[ \text{Li}_{2-r}\left(s^{-w} e^{-i\lambda_I - N^\alpha |\rho(x)|}\right) - (-1)^r \text{Li}_{2-r}\left(s^{w} e^{i\lambda_I - N^\alpha |\rho(x)|}\right)\Big].
\end{align}
The details of this large-$N$ first and small-$\beta$ expansions can found in appendix \ref{AppNLocal}. Note that if we consider a bifundamental representation of $U(N) \times U(N)$, we have the leading result in $N$, exact in $\beta$: \begin{align}
   \log Z_{\text{1-loop},I}^{(N,\bar{N})} &=  -\frac{N^\alpha}{2\beta}(i \lambda_I + \beta(1 - r_I))\sum_{a,b}|x_a - \tilde{x}_b| \\
     &- 2\frac{ N^{2 - \alpha}}{2\beta} \sum_{r=0}^3 \frac{(2 \beta)^r}{r!} B_r\left(\frac{r_I}{2}\right) \int dx \rho(x) \tilde{\rho}(x) P_{(3-r)}(\tilde{y}(x) - y(x) - \lambda_I).
\end{align}
Exactness means that as an expansion in small $\beta$, the full answer contains only terms proportional to $1/\beta$ and $\beta^{i=0,1,2}$. To simplify this, we use the reflection symmetry of the polylogarithms quoted in equation (\ref{eq:Pna}).
Similarly, for an antifundamental representation: \begin{align}
   \log Z_{\text{1-loop},I}^{(\bar{N},N)} &=  -\frac{N^\alpha}{2\beta}(i \lambda_I + \beta(1 - r_I))\sum_{a,b}|\tilde{x}_a - x_b| \\
     &- 2\frac{ N^{2 - \alpha}}{2\beta} \sum_{r=0}^3 \frac{(2 \beta)^r}{r!} B_r\left(\frac{r_I}{2}\right) \int dx \rho(x) \tilde{\rho}(x) P_{(3-r)}(y(x) - \tilde{y}(x) - \lambda_I).
\end{align}
Considering the earlier expressions, we can see that at a Lagrangian level, all of the contributions truncate at order $\beta^2$, so that we could likewise construct a local Lagrangian after determining the relation between $\rho(x)$ and $\rho^*(x)$ imposed at the $O(N^{2+\alpha})$ level. That is, from the above rules we may write 
\begin{align}
Z \sim \int \mathcal{D} \rho \mathcal{D} \delta y e^{-S[\rho(x), \delta y(x)]}, \quad S[\rho(x), \delta y(x)] = \frac{N^{3/2}}{\beta}\int \mathcal{L}[x, \rho(x), \delta y(x)] dx + o(N^{3/2}),
\end{align}
where the Lagrangian has been appropriately assembled with the above rules. 

From here, the Lagrangian could be extremized and a further Legendre transformation would lead to the entropy of the dual AdS$_4$ black hole in the full Cardy-like expansion, as we have considered in the ABJM case. 

We emphasize that keeping $N$ finite does not seem to lead to a truncated expression in powers of $\beta$ for the SCI (see appendix   \ref{qPochDev} for further details). This situation indicates an obstruction for a standard EFT interpretation at finite $N$ which is quite different from the situation for the 4d SCI \cite{Cassani:2021fyv}. It would be interesting to explore this situation, possibly in the non-supersymmetric setup along the lines of \cite{Bhattacharyya:2007vs,Shaghoulian:2015kta}. 

It is worth highlighting some of the similarities and differences between our approach here and relevant results in the literature. First, we compare with the discussion in \cite{Choi:2019zpz} where a factorization in terms of a holomorphic combination of holonomies and color fluxes was advanced. Note that here we do not postulate such factorization. In fact, that precise factorization seems to hold only at the leading order in $N$ and in $\beta$, that is, only for the term proportional to $1/\beta$. Recently, a discussion of the  refined topologically twisted index has been presented in \cite{Hosseini:2022vho}, a central tool in this approach has been the holomorphic block factorization established in \cite{Beem:2012mb}. The analysis in \cite{Hosseini:2022vho} is leading in $N$ and includes higher order corrections in $\beta$ that appear as a result of certain shifts in the chemical potentials; there does not seem to exist a truncation in the $\beta$ expansion for the corresponding Lagrangian. Although we are discussing a different observable, the superconformal index, we see various analogies in the shifts of chemical potentials that we implement. It would be interesting to explore whether our approach leads to an exact truncation in the $\beta$ expansion for the Lagrangian associated with the refined topologically twisted index.

\section{Corrections to the black hole entropy}\label{Sec:Entropy}
In the preceding section we systematically studied, in the large-$N$ limit, sub-leading corrections in $\beta$ for the SCI. The SCI is defined for fixed values of the chemical potentials, that is, in the grand-canonical ensemble. In this section we study aspects of the corresponding entropy which constitute a prediction for the black hole entropy on the dual supergravity side. Given that the entropy is defined in the micro-canonical ensemble, we should implement a change of ensemble via an inverse Laplace transform. For large charges the Laplace transform can subsequently be approximated by a saddle point evaluation. The leading result, that is, the $1/\beta$ term, was successfully matched to the Bekenstein-Hawking entropy of the dual black hole \cite{Choi:2019zpz, Nian:2019pxj}. In the current analysis we will also include logarithmic corrections to the entropy that arise as the result of Gaussian determinants around the saddle point solution. Such treatment has recently been performed in \cite{David:2021qaa} to determine logarithmic in $N$  corrections for the entropy of AdS$_5$ black holes and black strings. Since our evaluation has been limited to the large-$N$ limit, our search for logarithmic corrections pertains to logarithmic  corrections in $\beta$  as we depart from the strict Cardy  limit, $\beta\to 0$. In section \ref{subsec:logEnsemble}, given that there were no logarithmic in $\beta$ terms in our treatment of the SCI,  we shall discuss the logarithmic correction in $\beta$ associated to changing  of ensembles. 
 
In section \ref{subsec:Ent} we analyze the corrections to the black hole entropy due to our evaluation of the SCI beyond the strict Cardy limit. We find, somewhat unexpectedly, that the functional form of the entropy is preserved in terms of charges that are non-trivially shifted by $\beta$-dependent quantities. There has been a number of discussions regarding the nature of certain non-linear constraint among the charges. It was originally suggested in \cite{Chong:2005hr,Cvetic:2005zi} that the non-linear constraint on the charges is a requirement to avoid the presence of closed time-like curves on the supergravity solutions. From the field theory point of view, the constraint was motivated as a requirement to guarantee that the entropy remains a real quantity  \cite{Cabo-Bizet:2018ehj,Choi:2018hmj,Benini:2018ywd}. A more algebraic justification for the non-linear constraint was advanced in \cite{Larsen:2021wnu} in the context of AdS$_3$ black holes.

Our treatment naturally leads to a non-linear constraint among the charges, albeit for the $\beta$-shifted ones. Since we are working in the field theory context, our constraint also arises from the requirement of having a real entropy. It is worth noting that when we refer to the $\beta$-shifted charges we mean to keep $\beta$ as a place holder  which is, in fact, determined in terms of the charges. Clearly, the final micro-canonical expression for the entropy should be purely in terms of the charges. However, for the purpose of keeping track of the effect of the  $\beta$ expansion,  it will suffice to understand general behaviors of the shifted charges rather than a full closed expression. It will be interesting to explore this problem further and obtain the full expression entirely and explicitly in terms of charges. In particular, such expression would give us more freedom to relax the reality condition, thus accessing the imaginary phase of the microcanonical SCI.
With some abuse of terminology, one can think of this as the entropy picking up an imaginary part. Such a quantity has been interpreted in the numerical treatment of the microcanonical SCI of  ${\cal N}=4$ SYM  in \cite{Agarwal:2020zwm} (see also \cite{Murthy:2022tbj}). In this framework, the real part of the micro-canonical SCI defines the saddle point while, the imaginary part, determines certain frequency of oscillations. In view of this interpretation it would be interesting to explore more general regimes of charges that allow us to find a signature of the $\beta$ corrections in the imaginary part of the entropy.

We first present a general discussion and then particularize to the case of the SCI of ABJM theory.

\subsection{Logarithmic in $\beta$ correction from change of ensemble} \label{subsec:logEnsemble}

Let us generically denote by $Z_{\text{GC}}$ the partition function evaluated in the grand-canonical ensemble and by $Z_{\text{MC}}$  the partition function in the microcanonical ensemble,  i.e.,  computed for fixed values of the charges.

We consider $D$ chemical potentials $\lambda_{I}$ ($i=1, \cdots, D-1$) satisfying the constraint,
\begin{align}
\sum_{I=1}^{D-1} c_I\lambda_I+c_D \beta & =  n_0, \label{eq:constraint}
\end{align}
where $c_I $ are constants and we have defined $\mu_{D} =\beta$. Recall that we are interested in studying the Cardy-like expansion, the small $|\beta|$ regime. We implement the inverse Laplace transform which takes us from the grand-canonical ensemble to the microcanonical ensemble

\begin{align}
Z_{\text{MC}}(Q, J) & = \int d^{D-1} d \beta \lambda\, d \Lambda \exp \left[ \log Z_{\text {GC}}+ \sum_{I=1}^{D-1}  Q_I \lambda_I + \beta J + \Lambda\left(\sum_{I=1}^{D-1} c_I \lambda_I +c_D\beta - n_0\right)\right],
\end{align}
where $\Lambda$ is the Lagrange multiplier enforcing  the constraint \eqref{eq:constraint}. 
We know the partition function in the grand-canonical ensemble takes the form
\begin{align} \label{eq:GC result}
    \log Z_{\text{GC}}\Big{|}_{\text{leading}} = S_E(\lambda, \beta),
\end{align} 
which is homogeneous of degree one in $\lambda, \beta$ and it scales as $\beta^{-1}$. 
Let us focus on the leading regime \eqref{eq:GC result} and study the subleading logarithmic contribution in $\beta$ associated to the change of ensemble. We then write:
\begin{align}
Z_{\text{MC}}(Q, J) & =  \int d^{D-1} d \beta \lambda\, d \Lambda \exp \left[ S_E(\lambda, \beta)+ \sum_{I=1}^{D-1}  Q_I \lambda_I + J\beta  + \Lambda\left(\sum_{I=1}^{D-1} c_I \lambda_I +c_D\beta - n_0\right)\right]. \label{eq:Laplace}
\end{align}
For large values of the charges we can approximate the integral \eqref{eq:Laplace} using the saddle point method which yields to the following equations:  
\begin{align}
    \begin{split}
    \frac{\partial S_E(\lambda, \beta)}{\partial \lambda_I}  & =-( Q_I + c_I \Lambda ), \label{eq:saddle0}\\
    \frac{\partial S_E(\lambda, \beta)}{\partial \beta}& = - (J +c_D \Lambda),\\
\sum_{I=1}^{D-1} c_I \lambda_I +c_D\beta  &=  n_0.
    \end{split}
\end{align}
For compactness, it will useful to denote $J\equiv Q_D, \quad \beta \equiv \lambda_D$.
The homogeneity of $S_E(\lambda, \beta)$ implies the following crucial relation
\begin{align} \label{eq: homogeneity of I}
S_E(\lambda, \beta) & = \sum_{I=1}^{D-1}  \lambda_I \frac{\partial S_E(\lambda, \beta)}{\partial \lambda_I}+ \beta \frac{\partial S_E(\lambda, \beta)}{\partial \beta}.
\end{align}

Evaluating at the saddle point values, we obtain
\begin{align}
S_E(\lambda^*, \beta^*) & =- \sum_{I=1}^{D-1}  \lambda_I^{\star} \left(Q_I + c_I \Lambda\right)- \beta^* (J+ c_D \Lambda),
\end{align}
such that the saddle point imposed on \eqref{eq:Laplace} yields
\begin{align}
\begin{split}
Z_{\text{MC}} & \approx  \exp \left\{- \sum_{I=1}^D  \lambda_I^{\star} \left(Q_I + c_I \Lambda\right) + \sum_{I=1}^D  Q_I \lambda^{\star}_I+ \Lambda\left(\sum_{I=1}^D c_I \lambda^{\star}_I - n_0\right) -\frac{1}{2} \log \det \left(H\right)\right\}\\
& = e^{n_0 \Lambda -\frac{1}{2} \log \det \left(H\right)}.
\end{split}
\end{align}

The Hessian $H$ has the form
\begin{align}
 H &  = \begin{pmatrix}
\frac{\partial^2 S_E}{\partial \lambda_1^2} && \cdots && \frac{\partial^2 S_E}{\partial \lambda_1 \partial \beta}  &&\frac{\partial^2  S_E}{\partial \lambda_1 \partial \Lambda} \\
. && . && . && . \\
. && . && . && . \\
\frac{\partial^2  S_E}{\partial \beta \partial \lambda_1} && \cdots && \frac{\partial^2 S_E}{\partial \beta^2} && \frac{\partial^2  S_E}{\partial \beta \partial \Lambda}  \\
\frac{\partial^2  S_E}{ \partial \Lambda\partial \lambda_1}  && \cdots && \frac{\partial^2  S_E}{\partial \Lambda \partial \beta}  && \frac{\partial^2 S_E}{\partial \Lambda^2 }  \\
\end{pmatrix}.
\end{align}

The scaling of $S_E(\lambda, \beta)$ as $\beta^{-1}$ implies the following scaling of $H$
\begin{align}
\det H & \sim \det \begin{pmatrix}
\mathcal{O}(\beta^{-1}) && \cdots && \mathcal{O}(\beta^{-2})  && c_1\\
. && . && . && . \\
. && . && . && . \\
\mathcal{O}(\beta^{-2}) && \cdots && \mathcal{O}(\beta^{-3}) && c_D \\
c_1 && \cdots && c_D && 0 \\
\end{pmatrix} \sim \mathcal{O}(\beta^{-2(D-2)}).
\end{align}
Therefore,  the logarithmic correction in $\beta$ associated to the change of ensemble is:
\begin{align}
    \log Z_{\text{MC}} & = n_0 \Lambda + (D-2) \log \beta.
\end{align}
Our interest in the logarithmic in $\beta$ corrections arises from one of its potential interpretations. Namely, it is plausible to interpret the coefficient of $\log(\beta)$  as a sort of two-dimensional anomaly. Indeed, given $\log(\beta)=\log(R_{S^1}/R_{S^2}) =-\log (R_{S^2}/\epsilon)$, one can think of $R_{S^1}$ as a uv cutoff for the effective two-dimensional theory on $S^2$ of radius $R_{S^2}$. It would be interesting to further pursue this interpretation in the framework\footnote{We thank Lorenzo Di Prietro for insightful discussions on this topic.} of EFT. We have ignored $\log N$  corrections because we worked exclusively in the strict large $N$ limit. However, those $\log N$, contributions contain valuable physical information and would be worth determining. 
  \subsection{Generic structure of Cardy-like corrections} \label{subsec:Ent}
  So far we have dealt with the entropy function evaluated at leading order, that is, keeping only the $\mathcal{O}(\beta^{-1})$ in its small $|\beta|$ expansion. We have been able to probe the logarithmic $\beta$ correction associated to the change of ensemble and showed that it is strictly controlled by the number of independent chemical potentials in the theory. Now we would like to move on and study the possibility of including higher  order corrections in $\beta$ and study how they modify the micro-canonical partition function that should yield  the entropy of the dual black hole with electric charges $Q_I$ and angular momentum $J$. In this discussion we will assume some basic properties of the corrected entropy function and we shall explore their implications in the process  of extracting the entropy. The properties we demand to be preserved upon inclusion of $\beta$-corrections are the following:
  
  \begin{itemize}
      \item[i)] The homogeneity property of the entropy function is preserved once all perturbative corrections in $\beta$ have been included. Namely, both $\log Z_{\text{GC}}(\lambda, \beta)$ and $S_E(\lambda, \beta)$ are homogeneous of degree one.

      \item[ii)] There are no further logaritmic corrections in $\beta$ to the grand-canonical partition function. 
  \end{itemize}
  
The above set of properties is sufficient to determine the entropy, including its modifications once higher order $\beta$-corrections are incorporated. Let us then write:
  \begin{align}
      \log Z_{\text{GC}}(\lambda, \beta) & = S_E(\lambda, \beta) \mathcal{K}(\lambda, \beta),
  \end{align}
where $\mathcal{K}(\lambda, \beta)$ encodes the  corrections and it has to reduce to $1$ as $\beta \to 0$ by construction\footnote{Note that we can always find such $\mathcal{K}$ since it is just the ratio  $\frac{ Z_{\text{GC}}(\lambda, \beta) }{S_E(\lambda, \beta)}$}. 

By virtue of $i)$ we can prove that $\mathcal{K}(\lambda, \beta)$ must satisfy 
\begin{align}
    \sum_{I=1}^{D-1} \lambda_I \frac{\partial \mathcal{K}(\lambda, \beta)}{\partial \lambda_I}+ \beta \frac{\partial \mathcal{K}(\lambda, \beta)}{\partial \beta} =0 \label{eq:propK},
\end{align}
which follows from Euler theorem on homogeneous functions of degree one applied to both $\log Z_{\text{GC}}(\lambda, \beta)$ and $S_E(\lambda, \beta)$. In fact, there are infinitely many possible solutions to the partial differential equation \eqref{eq:propK}, and they are given as:
\begin{align}
    \mathcal{K}(\lambda, \beta) & = \mathcal{K}\left(\frac{\lambda_l}{\lambda_1}, \frac{\beta}{\lambda_1}\right), \quad l=2, \cdots, D-1. \label{eq:Kform}
\end{align}
To implement the saddle point approximation in the inverse  Laplace transformation we need to solve:
\begin{align}
    \begin{split}
        S_E(\lambda, \beta) \frac{\partial \mathcal{K}(\lambda, \beta)}{\partial \lambda_I} + \mathcal{K}(\lambda, \beta)\frac{\partial S_E(\lambda, \beta)}{\partial \lambda_I} + Q_I+ c_I\Lambda  & =0, \quad I =1,\cdots, D-1 \label{eq:correctedsaddlesS0}\\
         S_E(\lambda, \beta) \frac{\partial \mathcal{K}(\lambda, \beta)}{\partial \beta} + \mathcal{K}(\lambda, \beta)\frac{\partial S_E(\lambda, \beta)}{\partial \beta} + J+ c_D \Lambda   & =0.
    \end{split}
\end{align}

Rearranging \eqref{eq:correctedsaddlesS0} we can write:
\begin{align}
    \begin{split}
       \frac{\partial S_E(\lambda, \beta)}{\partial \lambda_I} & = - \frac{1}{\mathcal{K}(\lambda, \beta)} \left(Q_I+ S_E(\lambda, \beta) \frac{\partial \mathcal{K}(\lambda, \beta)}{\partial \lambda_I}+ c_I \Lambda \right), \quad I =1,\cdots, D-1 \label{eq:neweq0} \\
       \frac{\partial S_E(\lambda, \beta)}{\partial \beta} & =- \frac{1}{\mathcal{K}(\lambda, \beta)} \left(J +  S_E(\lambda, \beta) \frac{\partial \mathcal{K} (\lambda, \beta)}{\partial \beta}+ c_D \Lambda \right)
    \end{split}
\end{align}

Let us define:
 \begin{align}
     \label{eq:redef0}
     \widetilde{Q}_I & = \frac{1}{\mathcal{K}(\lambda, \beta)}\left(Q_I +S_E(\lambda, \beta)\frac{\partial \mathcal{K}(\lambda, \beta)}{\partial \lambda_I}\right), \quad I =1,\cdots,D-1, \\
     \widetilde{J} & =  \frac{1}{\mathcal{K}(\lambda, \beta)} \left(J +S_E(\lambda, \beta) \frac{\partial \mathcal{K} (\lambda, \beta)}{\partial \beta} \right), \\
     \widetilde{\Lambda} & = \frac{\Lambda}{\mathcal{K}(\lambda,\beta)}.
 \end{align}
 
We have that the derivatives of the leading entropy function have the same structure as they had in the strict cardy-like limit if we replace $Q_I \rightarrow \widetilde{Q}_I, \, J \rightarrow \widetilde{J}, \,\Lambda \rightarrow \widetilde{\Lambda}$. We then have:
\begin{align}
     \begin{split}
        \frac{\partial S_E(\lambda, \beta)}{\partial \lambda_I}   & =-(\widetilde{Q}_I+ c_I \widetilde{\Lambda}), \quad I =1,\cdots, D-1, \label{eq:leadingsaddlesS0}\\
        \frac{\partial S_E(\lambda, \beta)}{\partial \beta}   & =-(\widetilde{J}+ c_D \widetilde{\Lambda}).
    \end{split}
\end{align}
Recall that we need to solve for $\Lambda$ since it encodes the black hole entropy, therefore, we need to find a combination of the derivatives of $S_E(\lambda, \beta)$ that can be expressed purely in terms of $Q_I, \,J$ and $\Lambda$. This is the prescription followed in the strict Cardy-like limit in \cite{Choi:2019zpz,Choi:2018fdc,Cassani:2019mms}. By writing \eqref{eq:leadingsaddlesS0} we have reduced the problem of finding the corrected entropy to the same problem solved at leading order, only this time in terms of shifted charges \eqref{eq:redef0}. This simple fact already imposes non-trivial constraints on the form of $\mathcal{K}(\lambda, \beta)$ and its derivatives, since we want that the shifts of the charges reduce to zero in the $\beta \rightarrow 0$ limit. To be concrete, we demand that:
\begin{align} \label{eq:deriv}
\begin{split}
  S_E(\lambda, \beta)\frac{\partial \mathcal{K}(\lambda, \beta)}{\partial \lambda_I}\Bigg{|}_{\beta \rightarrow 0} & =0, \quad I =1,\cdots,D-1,  \\
  S_E(\lambda, \beta)\frac{\partial \mathcal{K}(\lambda, \beta)}{\partial \beta}\Bigg{|}_{\beta \rightarrow 0} & = 0.
  \end{split}
\end{align}
The second condition should be analyzed more carefully, since it will be modified after shifting the chemical potentials as: $\lambda_I \rightarrow \lambda_I + \alpha_I \beta$, with $\alpha_I$ are such that $|\alpha_I| \sim 1, \, I=1, \cdots, D-1$. This can be seen at the level of the Laplace transform \eqref{eq:Laplace}, where the above mentioned shifting can be absorbed into a redefinition of the angular momentum:
\begin{align}
    J \rightarrow J + \sum_{I=1}^{D-1}  \alpha_I Q_I
\end{align}
We then have that the second condition in \eqref{eq:deriv} should be modified as:
\begin{align} \label{eq:balance}
    S_E(\lambda, \beta)\frac{\partial \mathcal{K}(\lambda, \beta)}{\partial \beta}\Bigg{|}_{\beta \rightarrow 0} = -\sum_{I=1}^{D-1}\alpha_I Q_I.
\end{align}
Equation \eqref{eq:balance} will be crucial for the evaluation the entropy in the context of the ABJM SCI.

\subsection{The $\beta$-corrected entropy of ABJM}
To obtain the entropy we need to perform an inverse Laplace transformation which reduces to a Legendre transformation in the large-$N$ limit. Originally, we have the chemical potentials $\lambda_I$ in terms of which we write:
\begin{align}
   \mathcal{A}(\lambda , \beta) & =     \log Z_{ABJM}(\lambda , \beta) + i \sum_{I =1}^4 Q_I \lambda_I -\beta J + i  \Lambda \left(\sum_{I =1}^4 \lambda_I- 2\pi\right),
\end{align}
where the function $\mathcal{A}(\lambda, \beta)$ is the one we need to extremize in order to find the entropy. The coefficients of the chemical potentials have been chosen in accordance to \eqref{eq:fug}. As we saw in previous sections, it will be convenient to shift the chemical potentials as $\lambda_I = \Delta_I - i\frac{\beta}{2}$, which yields:
\begin{align}
    \mathcal{A}(\Delta, \beta) & =  \log Z_{ABJM}(\Delta, \beta)  +i  \sum_{I =1}^4 Q_I \Delta_I -\beta \tilde{J} +i \Lambda \left(\sum_{I =1}^4 \Delta_I- 2\pi - i \beta\right),\\
\tilde{J}    & = J - \frac{1}{2}\sum_{I=1}^4 Q_I.
\end{align}
Note  that we have a arrived to a situation as the one described at the end of subsection \ref{subsec:Ent} with $\alpha_I = \frac{1}{2}, \, \forall  \, I=1, \cdots, 4$. \par 
Collecting various results from subsection \ref{Sec:SPApproximation}, if we set $k=1$ in \eqref{eq:Z-ABJM-Final}, we have the following expression for the SCI of ABJM  including subleading corrections in $\beta$:
\begin{align}
 \log Z_{ABJM}(\Delta, \beta) &=-\frac{i N^{3/2} \sqrt{\frac{2 }{3}}  \sqrt{\frac{ \Delta _1 \Delta _3 \left(4 \pi  \beta ^2+3 \left(\Delta _1+\Delta _3\right) \left(\Delta _2+\Delta _3\right) \Delta _4\right) \left(4 \pi  \beta ^2+3 \Delta _2 \left(\Delta _1+\Delta _3\right) \left(\Delta _1+\Delta _4\right)\right)}{\left(\Delta _1+\Delta _3\right)  \left(4 \pi  \beta ^2+3 \left(\Delta _1+\Delta _3\right) \left(\Delta _2+\Delta _3\right) \left(\Delta _1+\Delta _4\right)\right)}}}{3 \beta }.
\end{align}
 Since we can always rewrite the result by making the replacement $\pi \rightarrow \frac{1}{2}(\sum_{I=1}^4 \Delta_I- i \beta)$ then the function $\log Z_{ABJM}$ is homogeneous of degree one. We will exploit this property even though sometimes we leave the $\pi$ in the formula for the sake of compactness.
 If we denote by  $S_{\text{E}}$ the leading result in the Cardy-like limit, then we can rewrite $\log Z_{ABJM}$ as follows:

\begin{align}
\begin{split}
    \log Z_{ABJM}(\Delta, \beta) & = S_{E}(\Delta, \beta) \mathcal{K}(\Delta, \beta),\label{eq:LZABJM}\\
  S_E(\Delta, \beta)  & =-\frac{i \sqrt{2 }N^{3/2}  \sqrt{\Delta _1 \Delta _2 \Delta _3 \Delta _4}}{3 \beta }, \\
    \mathcal{K}(\Delta, \beta) & = \frac{\left(\frac{4 \pi \beta^2}{3(\Delta_1 + \Delta_3)\Delta_2 \Delta_4}+1+ \frac{\Delta_1}{\Delta_4}\right)\left(\frac{4 \pi \beta^2}{3(\Delta_1 + \Delta_3)\Delta_2 \Delta_4}+1+ \frac{\Delta_3}{\Delta_2}\right)}{\frac{4 \pi \beta^2}{3(\Delta_1 + \Delta_3)\Delta_2 \Delta_4}+\left(1+ \frac{\Delta_1}{\Delta_4}\right)\left(1+ \frac{\Delta_3}{\Delta_2}\right)}.
    \end{split}
\end{align}
Note that, as expected, $\mathcal{K}(\Delta, 0) =1$. 
It is worth pointing out that in the regime where $\Delta_1=\Delta_3, \, \Delta_2 =\Delta_4$, the function $\mathcal{K}(\Delta, \beta)$ simplifies considerably:
\begin{align*}
    \mathcal{K}(\Delta,\beta)\Big{|}_{\Delta_1=\Delta_3, \, \Delta_2 =\Delta_4} & =\frac{\left(2 \pi  \beta ^2+3 \Delta _1 \Delta _2^2+3 \Delta _1^2 \Delta _2\right){}^2}{3 \Delta _1 \Delta _2 \left(\Delta _1+\Delta _2\right) \left(\pi  \beta ^2+3 \Delta _1 \Delta _2^2+3 \Delta _1^2 \Delta _2\right)}.
\end{align*}

The saddle point equations take the form:
\begin{align}
     \begin{split}
        \frac{\partial S_E(\Delta, \beta)}{\partial \Delta_I}   & =- i (\widetilde{Q}_I+ \widetilde{\Lambda}), \quad I =1,\cdots, 4 \label{eq:leadingsaddlesS}\\
        \frac{\partial S_E(\Delta, \beta)}{\partial \beta}   & =-(\widetilde{J}- \widetilde{\Lambda}).
    \end{split}
\end{align}
Consider now the derivatives of $S_E$:
\begin{align}
         \frac{\partial S_E}{\partial \Delta_I} & = -\frac{i N^{3/2} \sqrt{\Delta _1 \Delta _2 \Delta _3 \Delta _4}}{3 \sqrt{2} \beta  \Delta _I} , \quad 
       \frac{\partial S_E}{\partial \beta}  =  \frac{i \sqrt{2} N^{3/2} \sqrt{\Delta _1 \Delta _2 \Delta _3 \Delta _4}}{3 \beta ^2}.
\end{align}
One can then verify that:
\begin{align}
   0 = \prod_{I=1}^4\left( \frac{\partial S_{E}}{\partial \Delta_I} \right)+ \frac{N^3}{72} \left(\frac{\partial S_{E}}{\partial \beta}\right)^2. \label{eq:gen}
\end{align}
This relation among the derivatives of $S_E$ crucially allowed to express $\Lambda$ purely in terms of charges and angular momentum, that is, the chemical potentials drop out of the equation. Concretely, \eqref{eq:gen} and \eqref{eq:leadingsaddlesS}
 yield a polynomial equation of degree $5$. The same line of reasoning goes through now, only that we have to use the corrected charges \eqref{eq:redef0}. Hence, we have:
\begin{align}
    \prod_{I=1}^4\left(\widetilde{Q}_I+\widetilde{\Lambda}\right) -\frac{N^3}{72}(\widetilde{J}- \widetilde{\Lambda} )^2 & = 0. \label{eq:5eq0}
\end{align}
which can be rewritten as:

\begin{align}
    \prod_{I=1}^4\left(\hat{Q}_I+\Lambda\right) -\frac{\hat{N}^{3}}{72}(\hat{J}- \Lambda )^2 & = 0. \label{eq:5eq}
\end{align}
where we have redefined:
\begin{align}
\begin{split}
    \hat{Q}_I & = Q_I - i S_E(\Delta,\beta) \frac{\partial \mathcal{K}(\Delta,\beta)}{\partial \Delta_I},  \quad I=1,\cdots, 4 \label{eq:hat}\\
    \hat{J} & =\widetilde{J}+ S_E(\Delta, \beta) \frac{\partial \mathcal{K}(\Delta,\beta)}{\partial \beta} \\
    \hat{N}^{3} & =N^3 \mathcal{K}(\Delta, \beta)^2,
    \end{split}
\end{align}
As emphasized in subsection \ref{subsec:Ent}, in order for the leading order result to be recovered from \eqref{eq:5eq}, we need that the derivatives of $\mathcal{K}(\Delta,\beta)$ vanish at $\beta = 0$. This property can be directly checked using the explicit form of $\mathcal{K}(\Delta, \beta)$ given in \eqref{eq:LZABJM}. In fact, now the conditions \eqref{eq:deriv} with the appropriate modification given as \eqref{eq:balance} read:
\begin{align} \label{eq:der1}
  S_E(\Delta, \beta)\frac{\partial \mathcal{K}(\Delta, \beta)}{\partial \Delta_I}\Bigg{|}_{\beta \rightarrow 0} & =0, \quad I =1,\cdots,4,  \\
  S_E(\Delta, \beta)\frac{\partial \mathcal{K}(\Delta, \beta)}{\partial \beta}\Bigg{|}_{\beta \rightarrow 0} & =\frac{1}{2}\sum_{I=1}^{4}Q_I. \label{eq:der2}
  \end{align}
The first $4$ relations in \eqref{eq:der1} are satisfied identically by  $S_E(\Delta, \beta)$ and $\mathcal{K}(\Delta, \beta)$ given in \eqref{eq:LZABJM}. The nontrivial information is now encoded in \eqref{eq:der2} since it provides a nontrivial relation between the critical values of $\Delta_I$ and $\beta$ and the charges $Q_I$. 

A particularly interesting regime is the one already studied in \cite{Choi:2018fdc} is the one corresponding to  $Q_1 =Q_3, \, Q_2 =Q_4$ (  $\Delta_1 =\Delta_3, \, \Delta_2 =\Delta_4$). In this regime, the condition \eqref{eq:der2} will allow to find  a series expansion of the corrected entropy in powers of $\beta$ with coefficients expressed fully in terms of charges. 

Let us then consider the simple situation $Q_1 =Q_3, \, Q_2 =Q_4$ in which case \eqref{eq:gen} yields:
\begin{align}
    9 (\hat{Q}_1+\hat{\Lambda})^2(\hat{Q}_1+\hat{\Lambda})^2= - 2 \hat{N}^3(\hat{J}-\hat{\Lambda})^2
\end{align}

that yields the following solution for $\Lambda$:
\begin{align}
  \Lambda & =- \frac{i}{3}\sqrt{\frac{
 9 \hat{Q}_1 \hat{Q}_2(\hat{Q}_1+\hat{Q}_2) -2 \hat{N}^{3} \hat{J}}{\hat{Q}_1 +\hat{Q}_2}} \label{eq:lambdaent0}
\end{align}
 Recall that $\Lambda$ is proportional to the black hole entropy. We thus find the main result of this section - a closed form for the entropy that includes all perturbative corrections in $\beta$ and takes the form of the leading expression but with shifted charges:
\begin{empheq}[box=\fbox]{equation}
\begin{split}
   \quad \quad \quad S& =  \frac{2 \pi}{3}\sqrt{\frac{9 \hat{Q}_1 \hat{Q}_2(\hat{Q}_1+\hat{Q}_2) -2 \hat{N}^{3} \hat{J}}{\hat{Q}_1 +\hat{Q}_2}}\\
  0 & = 2 \hat{N}^{3} \hat{J}^2 +  2 \hat{N}^{3} (\hat{Q}_1 + \hat{Q}_2) \hat{J} - 9 \hat{Q}_1 \hat{Q}_2 (\hat{Q}_1 +\hat{Q}_2)^2 . \label{eq:constraintreal} \quad \quad \quad 
  \end{split}
\end{empheq}
The explicit expressions for the charges are given by \eqref{eq:hat} and the constraint in the second line of \eqref{eq:constraintreal} guarantees the reality of the entropy  provided $\hat{Q}_I, \, \hat{J}$ are real \cite{Choi:2018fdc}. In general, the shifting of the charges may produce non trivial imaginary part. This seeming non-renormalization of the form of the entropy is reminiscent of a similar result for the topologically twisted index reported in \cite{Bobev:2022jte}, where the exact answer is obtained by certain shifts in the leading order expression.

 The value of $\beta = \beta^{*}$ that solves \eqref{eq:leadingsaddlesS} can be obtained in the simplified regime of two independent charges. Let us first try to find $\beta^*$ in the strict Cardy-like limit. For consistency, this critical value of $\beta$
should be such that it satisfies $|\beta| \ll 1$ for values of charges satisfying the constraint in \eqref{eq:constraintreal}. To implement this consistency check, let us analyze the critical values of chemical potentials in the strict Cardy-like limit. From equation \eqref{eq:leadingsaddlesS} at leading order:
\begin{align}
\begin{split}
    \frac{ \sqrt{2}N^{3/2} \Delta^*_2}{\beta^*} & =( Q_1 + \Lambda), \quad     \frac{ \sqrt{2}N^{3/2} \Delta^*_1}{\beta^*} =( Q_2 + \Lambda),\\
    \end{split}
\end{align}
which yields:
\begin{align}
    \begin{split}
     \Delta^*_1& = - \frac{3  \beta^*(Q_2+ \Lambda)}{\sqrt{2}N^{3/2}} , \quad  \Delta^*_2 = \frac{3  \beta^*(Q_1+ \Lambda)}{\sqrt{2}N^{3/2}}, \\
        \beta^* & = -\frac{2 i \pi N^{3/2}}{2N^{3/2} + 3 i \sqrt{2}(Q_1+Q_2+ 2 \Lambda)},\\
           \end{split}
\end{align}
where the value of $\Lambda$ is obtained from \eqref{eq:lambdaent0} evaluated at the leading order in $\beta$. In fact we find that there is a regime of charges that always leads to a value of $|\beta^*| \ll1$, hence being a sound perturbative parameter.  In terms of $N, Q_1, Q_2$ we have:
\begin{align}
   \beta^* &=  -\frac{2 i \pi  N^{3/2}}{2 N^{3/2}+ i\sqrt{2} \left(3(Q_1+ Q_2)-2 i  \sqrt{9Q_1Q_2+N^3-\sqrt{N^6+18 N^3Q_1Q_2}}\right)}
\end{align}

\begin{figure}[h!]
\centering
    \includegraphics[width=0.7\textwidth]{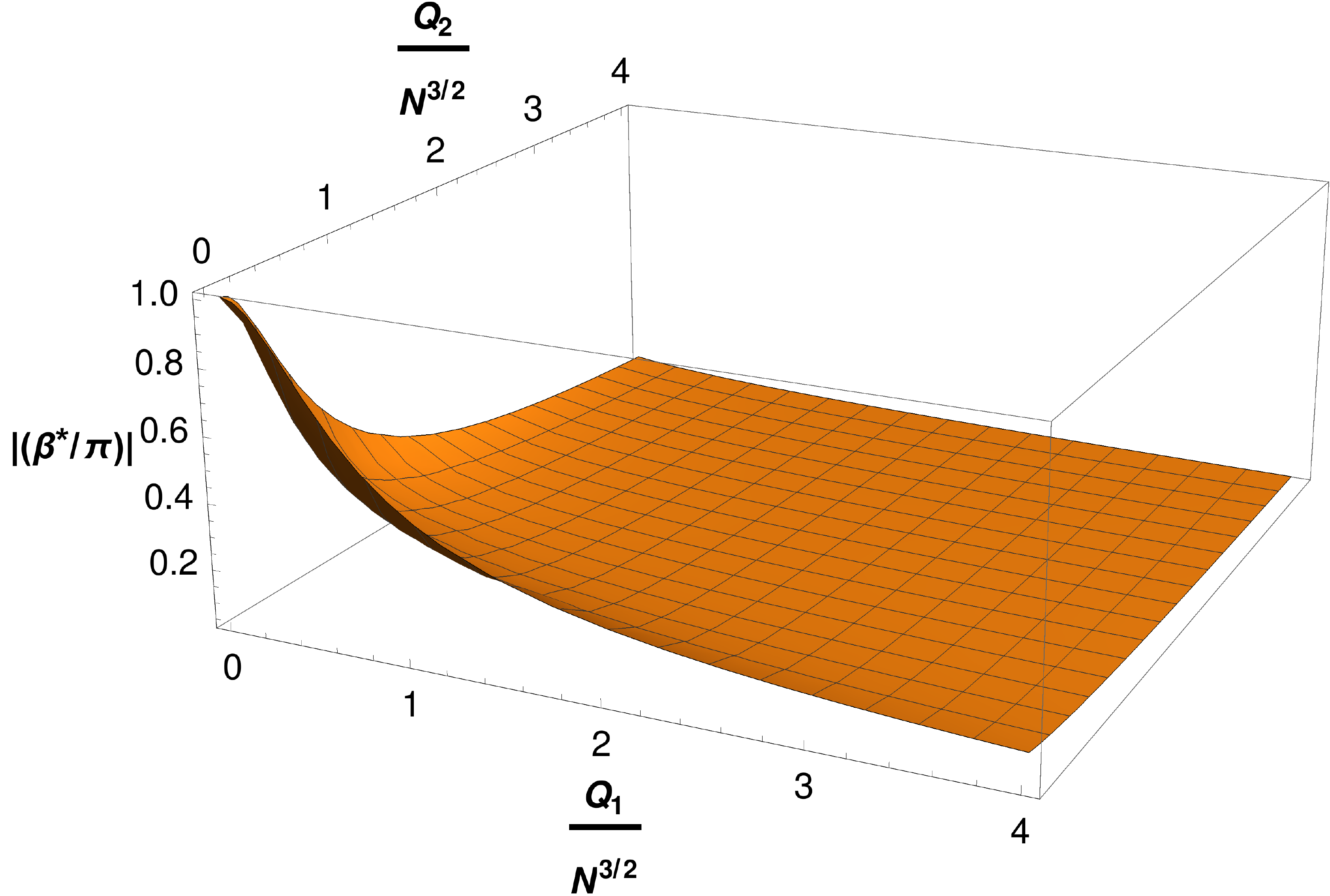}
    \caption{The figure shows the absolute values of $\beta^*/\pi$ as a function of the charges normalized by a $N^{3/2}$ factor. We observe a plateau for large positive values of charges  where $\beta^*/\pi$ becomes small and therefore can be consistently used as a perturbative parameter. Note that, for small values of the normalized charges, the critical value of $\beta$ approaches $- i \pi$, hinting to a potential relevance of the R-index. }\label{fig:betastar}
\end{figure}

In figure \ref{fig:betastar} we plot the absolute value of $\beta^*/\pi$ in terms of appropriately normalized charges.  For increasing positive values of the charges, we see that the critical values $|\beta^*/\pi| \ll1$ which is consistent with using $\beta$ as a perturbative parameter in the Cardy-like expansion. Note that, for small values of the normalized charges, the critical value of $\beta$ approaches $- i \pi$. Strictly speaking, we should interpret this as a breakdown of the perturbative regime in terms of $\beta$. Nevertheless, it is worth pointing out that this regime seems to single out a value of $\beta$ more suitable to be attained by using the $R$-charge index defined in Section \ref{Sec:SCI}.

\section{Conclusions}\label{Sec:Conclsions}

In this manuscript we have explored an explicit evaluation of the SCI of three-dimensional ${\cal N}=2$ supersymmetric field theories following its presentation via supersymmetric localization on $S^1\times S^2$. We have explicitly discussed the ABJM theory as an emblematic example. One significant goal we set was to study the index beyond the strict Cardy-limit, that is, to understand corrections in a small-$\beta$ expansion where $\beta$ is the ratio of the radius of $S^1$ to the radius of $S^2$. As one of our main results we found that, when taking the large-$N$ limit first, the Cardy expansion of the SCI truncates in the sense that it only contains terms proportional to $1/\beta$ and $\beta^{i=0,1,2}$. We have seen evidence (see appendix \ref{qPochDev}) that as we depart from the strict limit $\beta\to 0$, certain non-perturbative contributions of the form $e^{-1/\beta}$ arise. In equation (\ref{eq:Z-ABJM-Final}) we present a large-$N$ evaluation of the SCI for ABJM that is perturbatively exact in $\beta$. The generic behavior  for finite $\beta$ and finite $N$, could arguably be the culprit for the lack of a standard two-dimensional effective field theory interpretation. It is worth highlighting that the truncation in the small $\beta$ expansion takes place when we first take the large-$N$ limit. This circumstance is quite different from the four-dimensional SCI where the truncation takes places for any finite $N$. 

In section \ref{Sec:Entropy} we explore the implications of our evaluation of the index on the dual black hole entropy. We computed an interesting contribution of proportional to $\log \beta$ that arises from ensemble changing. Our main result in this direction  is given by equation (\ref{eq:constraintreal}) which is an expression for the entropy that is perturbatively exact in $\beta$ and takes the form of the leading order entropy but with the charges appropriately shifted. 

There are a number of interesting open questions that our work motivates. Arguably the most natural one relates to an EFT interpretation of the results along the lines of the EFT interpretation of the 4d SCI advanced in \cite{Cassani:2021fyv}. It is conceptually natural to think that if we consider a 3d theory on $S^1\times S^2$ and take the ratio of the $S^1$ radius to the radius of $S^2$ to be quite small, we expect that the resulting theory would organize itself as a two-dimensional theory on $S^2$. The fact that we obtained a sharp picture only after taking the large-$N$ limits obscures the EFT interpretation suggesting that an EFT interpretation is not readily present for finite $N$. At a technical level we can point to the crucial role, in determining the 3d SCI as an observable, that the non-perturbative contribution due to the sum over magnetic color fluxes plays. It would be interesting to elucidate this issue in more detail. One  approach to clarifying the situation in field theory would be to retreat to the supergravity setup where some progress has been achieved in the 4d context \cite{DiPietro:2014bca,Cassani:2021fyv}. We hope to report on this effort in the future. 

We have also explored how this small-$\beta$ corrections affect the entropy of the dual black hole. Interestingly, we have found that the corrected entropy preserves its leading order form once the charges are appropriately shifted; a similar result applies for the nonlinear constraint among the charges.  Preserving the leading form with shifted charges  is tantalizingly similar to a finite $N$ analysis of  ABJM where the exact expression was given by the appropriately shifted leading order expression \cite{Hristov:2022lcw}. It would be interesting to understand these corrections directly on the gravitational side.  The methods used in the paper could, in principle,  be expanded to determine the form of the logarithmic in $N$ corrections to the entropy. 

Another interesting direction to explore exploits the fact that the black hole entropy can be understood microscopically from the CFT$_3$ point of view, as we did here, but also via the Kerr/CFT$_2$ correspondence in the near-horizon region. It would be quite interesting to understand the role of the small-$\beta$  corrections from the point of view of the Kerr/CFT$_2$ and, more ambitiously, tackle the clarification of  Kerr/CFT$_2$ as embedded in the context of AdS$_4$/ABJM.

\section*{Acknowledgments}
We are grateful to Antonio Amariti, Arash Ardehali, Lorenzo Di Pietro, Seyed Morteza Hosseini, Imtak Jeon,  James T. Liu, Jun Nian, Sara Pasquetti, Augniva Ray, Alessia Segati, Christoph Uhlemann, Yu Xin and Alberto Zaffaroni for discussions.  This work is supported in part by the U.S. Department of Energy under grant DE-SC0007859. LPZ acknowledges support from an IBM Einstein Fellowship at the Institute for Advanced Study. AGL is supported by an appointment to the JRG Program at the APCTP through the Science and Technology Promotion Fund and Lottery Fund of the Korean Government and from the National Research Foundation of Korea(NRF) grant funded by the Korea government(MSIT) (No. 2021R1F1A1048531)

\appendix

\section{q-Pochhammer Identities}
\label{qPochDev}
Among 3d indices, the q-Pochhammer symbol is ubiquitous. In this section, we perform some manipulations towards a useful identity for analyzing their Cardy-like limit $\beta \rightarrow 0$. The main relation, which can be viewed as a slight generalization of a lemma due to Garoufalidis and Zagier \cite{Garoufalidis:2018qds}, reads:
\begin{align} \label{Eq:Pochhamer-IdentityMain}
(z q^a; q^2) = \exp \left( - \frac{1}{2 \beta} \sum_{r = 0} B_r\left(1 - \frac{a}{2}\right) \frac{(2 \beta)^r}{r!} \text{Li}_{2-r}(z)\right),
\end{align}
where $q=e^{-\beta}$. Unfortunately, this series in $r$ will always diverge for $\beta > 0$. In our analysis, this breakdown in the $\beta$-expansion underpins the need for a large-$N$ expansion which effectively truncates the $\beta$-expansion. 

To establish (\ref{Eq:Pochhamer-IdentityMain}), we take the logarithm of the q-Pochhammer symbol and apply its definition for $|z| < 1, q < 1$, and then perform a series of formal manipulations: \begin{align}
    \log (z q^a; q^2) &= \sum_{n=0}^\infty  \log \left(1 - z e^{- \beta(a + 2 n)}\right) \\
    &= -\sum_{n = 0}^\infty \sum_{k=1}^\infty\frac{1}{k} \left(z e^{- \beta(a + 2 n)}\right)^k \\
&= -\sum_{k=1}^\infty \frac{1}{k} z^k e^{- a \beta k} \sum_{n = 0}^\infty e^{- 2 \beta k n} \\
&= -\sum_{k=1}^\infty \frac{1}{k} z^k e^{- a \beta k} \frac{e^{2 \beta k}}{e^{2 \beta k} - 1}\\
&= -\sum_{k=1}^\infty \frac{1}{k} z^k \frac{1}{2 \beta k} \frac{(2 \beta k)e^{(1 - a/2) 2 \beta k}}{e^{2 \beta k} - 1}.
\end{align}
In the last line, the expression has been brought into the form of the generating functional for the Bernoulli Polynomials:\begin{align}
\frac{t e^{xt}}{e^t - 1} = \sum_{r = 0}^\infty B_r(x) \frac{t^r}{r!}. \label{eq:BernoulliGen}
\end{align}
Using (\ref{eq:BernoulliGen}) for $t = 2 \beta k$ and $x = 1 - \frac{a}{2}$ yields \begin{align}
   \log (z q^a; q^2) &= -\sum_{k=1}^\infty \frac{1}{k} z^k \frac{1}{2 \beta k} \sum_{r=0}^\infty B_r\left(1 - \frac{a}{2}\right)\frac{(2 \beta k)^r}{r!}\\
&= -\sum_{k=1}^\infty \frac{1}{k^{2 - r}} z^k \frac{1}{2 \beta} \sum_{r=0}^\infty B_r\left(1 - \frac{a}{2}\right)\frac{(2 \beta)^r}{r!}\\
&= -\frac{1}{2 \beta} \sum_{r=0}^\infty B_r\left(1 - \frac{a}{2}\right)\frac{(2 \beta)^r}{r!} \text{Li}_{2-r}(z).
\end{align}
We can then re-exponentiate both sides to recover the claimed (\ref{Eq:Pochhamer-IdentityMain}). 

To see that (\ref{Eq:Pochhamer-IdentityMain}) diverges for any $\beta>0$, and to get a sense for an appropriate cutoff in $r$, we first consider the summands, defined as \begin{align}
    c_r \equiv B_r\left(1 - \frac{a}{2}\right)\frac{(2 \beta)^r}{r!} \text{Li}_{2-r}(z).
\end{align} 
Due to the growth of the Bernoulli polynomials, the $c_r$ contributions for odd $r$ are dominant, so to perform a rough ratio test, we consider the ratio $c_{r+2}/c_r$ for odd $r$, and consider the leading terms for large $r$ and around $z \sim 1$: \begin{align}
    \frac{c_{r+2}}{c_r} &= \frac{B_{r+2}(1 - \frac{a}{2})}{B_{r}(1 - \frac{a}{2})} \frac{(2 \beta)^2}{(r + 1)(r+2)} \frac{\text{Li}_{-r}(z)}{\text{Li}_{2-r}(z)}\\
&\approx \frac{2+r}{r} \frac{B_{r+1}}{B_{r-1}} \frac{(2 \beta)^2}{(r + 1)(r+2)} \frac{r(r-1)}{(z-1)^2}\\
&\approx \frac{2+r}{r}  \left(\frac{r+1}{2 \pi e}\right)^2 \left(\frac{r+1}{r - 1}\right)^{r-1/2}  \frac{(2 \beta)^2}{(r + 1)(r+2)} \frac{r(r-1)}{(z-1)^2}\\
&\approx \left(\frac{\beta}{e \pi (z-1)}\right)^2 \frac{(r + 1)^{r + 1/2}}{(r - 1)^{r - 3/2}} \\
&\approx \left(\frac{\beta r}{\pi (z-1)}\right)^2.
\end{align}
Notice that for general $\beta$ and $z$, for sufficiently large $r$ this ratio will grow and the series will diverge. To establish a reasonable cutoff $R$, we should look for the largest value of $R$ such that the magnitude of this ratio is less than 1, so that \begin{align}
    c_{R+2}/c_{R} \approx \left(\frac{\beta R}{\pi t}\right)^2 = 1 \quad \rightarrow \quad R \sim \frac{|z-1| \pi}{\beta}. \label{eq:cutoffHueristic}
\end{align}

Despite the roughness of this analytic argument, the general lesson is born out numerically: The series will consistently diverge. Particularly, when $z$ is close to 1 and when $\beta$ is far from 0, the series diverges faster, and the ideal cutoff $R$ decreases. 

As a numerical example of this behavior, we set $a = \frac{1}{2}, z = e^{-2} \approx 0.135$ and plot the behavior and error of the partial sums:
\begin{align}
    P_{R}(\beta) &= -\frac{1}{2 \beta} \sum_{r = 0}^{R} B_r\left(1 - \frac{a}{2}\right) \frac{(2 \beta)^r}{r!} \text{Li}_{2-r}(z),
\end{align}
so that according to (\ref{Eq:Pochhamer-IdentityMain}), these partial sums would hope to approximate $\log(z q^a; q^2)$ and are simply truncations of the $\beta$-expansion of $\log(z q^a; q^2)$ at order $O(\beta^R)$:
\begin{align}
    \log(z q^a; q^2) &= P_R(\beta) + O(\beta^R).
\end{align}

\begin{figure}[t]
\centering
\includegraphics[width=0.7\textwidth]{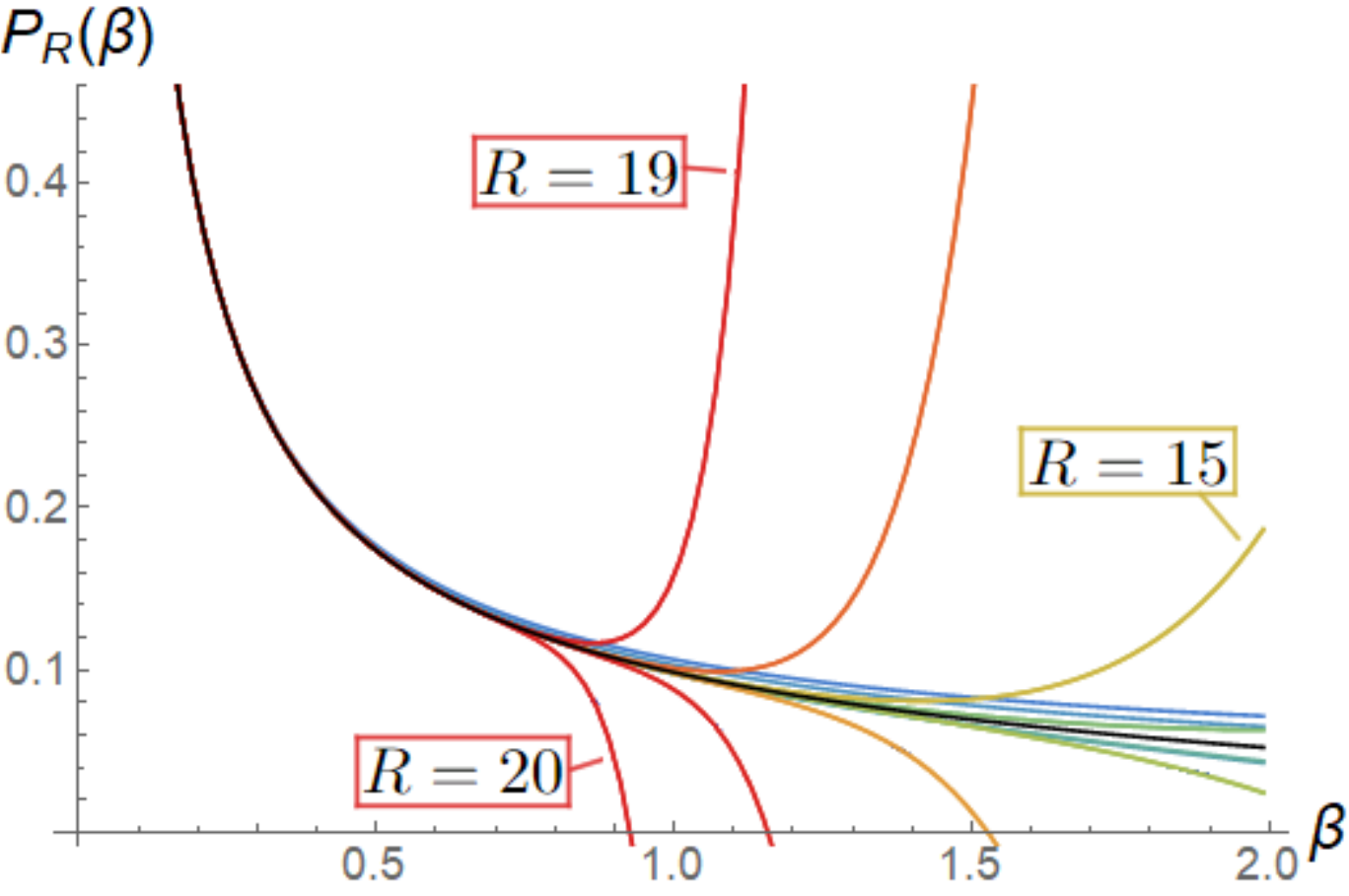}
\caption{ Successive approximations $P_{R}(\beta)$ to $\log(z q^a; q^2)$ are plotted for $z = e^{-2}, a = \frac{1}{2}$. Each $P_{R}(\beta)$ includes terms up to $O(\beta^R)$ in the perturbative expansion (\ref{Eq:Pochhamer-IdentityMain}). For clarity, $-\text{Re}[P_R(\beta)]$ and $-\text{Re}[\log(z q^a; q^2)]$ are plotted. }\label{fig:Pochhammer-pert-approx}
\end{figure}

These $P_R(\beta)$ are then plotted along with $\log(z q^a; q^2)$ in Figure \ref{fig:Pochhammer-pert-approx}. We can observe first that when $\beta$ is small the approximation can work well, but as either the cutoff $R$ or $\beta$ increase, there is greater deviation, as (\ref{eq:cutoffHueristic}) suggests. 

\begin{figure}[t]
\centering
\includegraphics[width=0.7\textwidth]{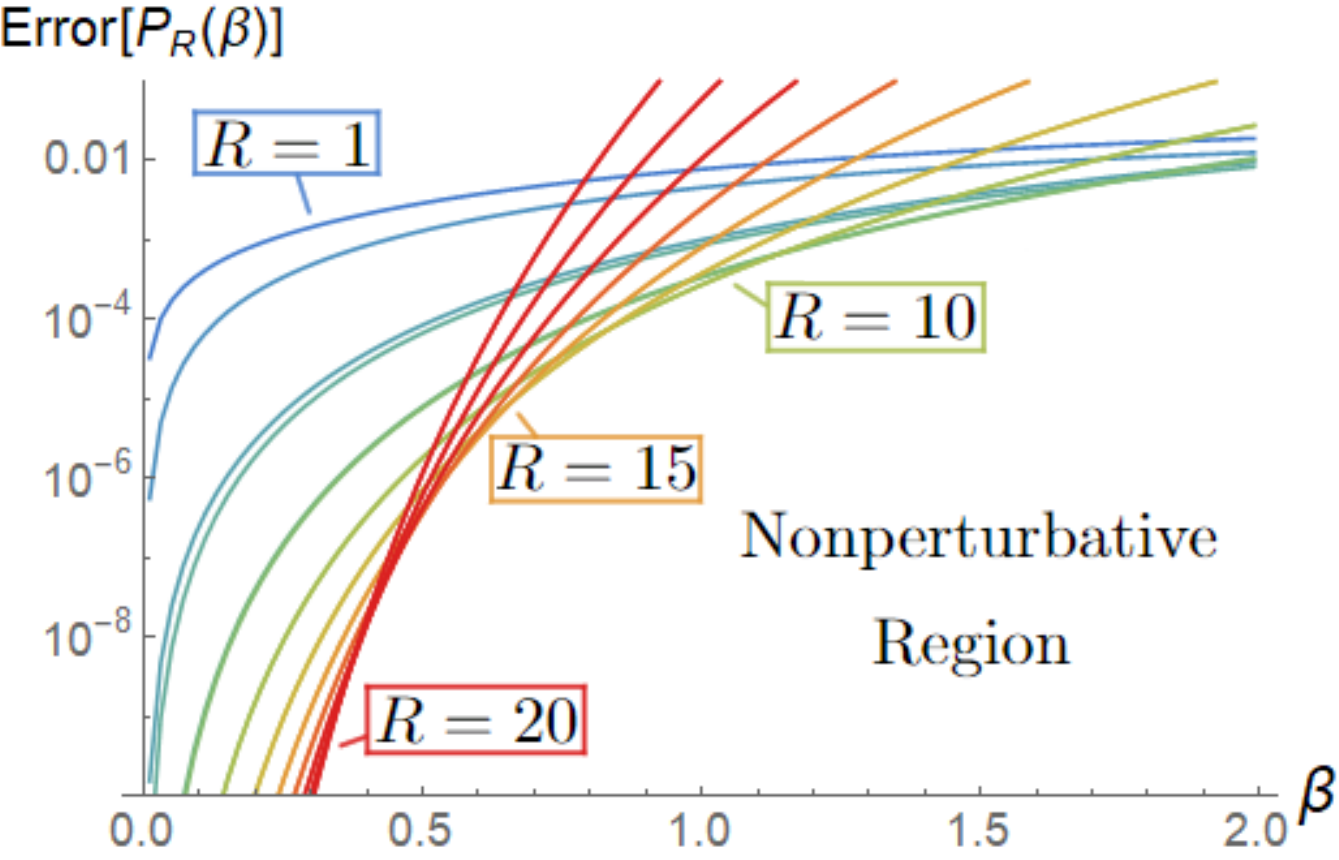}
\caption{The log plot of the difference between $\log(z q^a;q^2)$ and its truncated series expansions in $\beta$, $P_R(\beta)$. The bottom-right nonperturbative region shows that no cutoff will come within a certain range of the true answer.}\label{fig:Pochhammer-pert-error}
\end{figure}

To refine our understanding of the deviations, in Figure \ref{fig:Pochhammer-pert-error} we directly plot the error of each approximation, $|P_R(\beta) - \log(z q^a;q^2)|$. We can see the identity of the ideal cutoff shift as $R \sim \beta^{-1}$ and the presence of a persistent error in the $\beta$-expansion. This indicates a nonperturbative component of $\log(z q^a;q^2)$ which any $\beta$-expansion will fail to capture. 

From the expansion of the q-Pochhammer symbol, we can find a related identity:\begin{align}
    \sum_{n=0}^\infty \text{Li}_2 \left(z q^{a + 2n}\right) = \frac{1}{2\beta}\sum_{r=0}^\infty B_r\left(1 - \frac{a}{2}\right) \frac{(2 \beta)^r}{r!} \text{Li}_{3-r}\left(z\right). \label{eq:Li2-Identity}
\end{align}
For this, we use that
\begin{align}
\text{Li}_2(z) = \sum_{k=1}^\infty \frac{z^k}{k^2} = \sum_{k=1}^\infty\int_0^1 \frac{dp}{p} \frac{(p z)^k}{k}= \int_0^1 \frac{dp}{p} \text{Li}_1(p z).
\end{align}
So that \begin{align}
    \sum_{n=0}^\infty \text{Li}_2 \left(z q^{a + 2n}\right) &= \int_0^1 \frac{dp}{p} \sum_{n=0}^\infty \text{Li}_1 \left(p z q^{a + 2n}\right)\\
    &= - \int_0^1 \frac{dp}{p} \log (p z q^a; q^2) \\
    &=  \frac{1}{2 \beta} \sum_{r = 0} B_r\left(1 - \frac{a}{2}\right) \frac{(2 \beta)^r}{r!} \int_0^1 \frac{dp}{p}\text{Li}_{2-r}(p z)\\
    &=  \frac{1}{2 \beta} \sum_{r = 0} B_r\left(1 - \frac{a}{2}\right) \frac{(2 \beta)^r}{r!}\text{Li}_{3-r}(z)
\end{align}
where we have applied the identity for the expansion of the q-Pochhammer symbol.

\section{Details of large-$N$ limit} \label{AppNLocal}
Central to our analysis are certain ``localization arguments'' that we claim capture the large $N$ behavior of double sums that appear in the index. Here, we provide derivations of these results, while doing our best to control the errors in $N$ that arise, as well as provide some numerical evidence justifying the legitimacy of the manipulations. 

Specifically, the main result of this appendix is to find a large $N$ expansion of the following double sum: \begin{align}
    &S_{a,b} = \sum_{a=1}^N \sum_{b=1}^N \text{Li}_1\left(e^{i(h_1(x_a) + h_2(x_b)) - N^\alpha|x_{ab}|}\right) \label{eq:Li-1-DS} \\
    &= 2 N^{2 - \alpha} \int dx \rho(x)^2 \text{Li}_2\left(e^{i(h_1(x)+h_2(x))}\right) - \frac{N^{\alpha}}{6} \int dx \text{Li}_0\left(e^{i(h_1(x)+h_2(x))}\right) \\
    &- N^{2 - 2\alpha}\left(\rho(x_1)^2\text{Li}_3 \left(e^{i(h_1(x_1)+h_2(x_1))}\right)+\rho(x_N)^2\text{Li}_3 \left(e^{i(h_1(x_N)+h_2(x_N))}\right)\right).
\end{align}
In this section, we aim only to keep terms of order $O(N)$, although we recover a critical singularity at order $O(N^{1/2})$ which informs our saddle point analysis. 

\subsection{Manipulations}
To analyze this sum, we represent possible choices of $x_a$ with a non-decreasing function $x(a)$ defined so that $x(a) = x_a$ at each $a = 1,...,N$. The function $\rho(x)$ is then defined as \begin{align}
    \rho(x) = \frac{1}{N} \left(\frac{dx}{da}\right)^{-1}.
\end{align}
We will assume that as we take $N\rightarrow \infty$, the configurations $x_a$ that we consider will converge to a smooth function $\rho(x)$ which is independent of $N$ up to exponentially small corrections in $N$. Eq.~(\ref{eq:xa-choice}) gives an explicit example where this limit applies. 

To treat this situation, a common approach is to replace the sums present with integrals over the now continuum variable $x$. The key tool that enables us to do this is the Euler-Maclaurin formula, which contains boundary corrections to this prescription which turn out to be critical to our analysis. There are two related versions of this expansion, both of which we will make use of: \begin{align}
    \sum_{i=m+1}^{n} f(i)&=\int_{m+1}^{n} f(x) d x+\frac{f(n)+f(m+1)}{2}+\sum_{k=1}^{\infty} \frac{B_{2 k}}{(2 k) !}\left(f^{(2 k-1)}(n)-f^{(2 k-1)}(m+1)\right) \label{eq:EM-Same}\\
    \sum_{i=m+1}^{n} f(i)&=\int_{m}^{n} f(x) d x+\frac{f(n)-f(m)}{2}+\sum_{k=1}^{\infty} \frac{B_{2 k}}{(2 k) !}\left(f^{(2 k-1)}(n)-f^{(2 k-1)}(m)\right). \label{eq:EM-Extend}
\end{align}
Owing to the rapid growth of the Bernoulli numbers $B_{2k}$, this series will diverge. However, higher derivative terms are suppressed by powers of $N$, so in our large $N$ analysis, we will only consider the first derivative boundary corrections. 

To analyze the sum (\ref{eq:Li-1-DS}), we first decompose the polylogarithm into $k$-modes: \begin{align}
    \sum_{a=1}^N \sum_{b=1}^N \text{Li}_1\left(e^{i(h_1(x_a) + h_2(x_b)) - N^\alpha|x_{ab}|}\right) &= \sum_{a=1}^N \sum_{b=1}^N \sum_{k=0}^\infty \frac{1}{k} e^{ik(h_1(x_a) + h_2(x_b)) - kN^\alpha|x_{ab}|}\\
    &= \sum_{k=0}^\infty \frac{1}{k} \sum_{a=1}^N \sum_{b=1}^N f_k(a,b)
\end{align}
where \begin{align}
    f_k(a,b) = e^{ik(h_1(x_a) + h_2(x_b)) - kN^\alpha|x_{ab}|}
\end{align}
Our strategy will then be to first evaluate the double sums over the individual $f_k(a,b)$ and then at the end resum over $k$. 

Towards this, we first split the double sum into two parts: \begin{align}
    \sum_{a=1}^N \sum_{b=1}^N f_k(a,b) &= \sum_{a=1}^N \left[\frac{f_k(a,a)}{2} + \sum_{b = a+1}^N f_k(a,b)\right] + \sum_{a=1}^N\left[\frac{f_k(a,a)}{2} +  \sum_{b = 1}^{a-1} f_k(a,b)\right] \\
    &\equiv S_{k,a<b} + S_{k,a>b}.
\end{align}

We may then apply the Euler-Maclaurin formula (\ref{eq:EM-Extend}) to the inner sum over $b$ of $S_{k, a < b}$ to find \begin{align}
    S_{k, a < b} &= \sum_{a=1}^N \left[\frac{f_k(a,a)}{2} + \int_a^N db f_k(a,b) + \frac{f_k(a,N) - f_k(a,a)}{2} + \frac{1}{12} \left(\partial_2 f_k(a,N) - \partial_2 f_k(a,a)\right)\right] \\
    &= \sum_{a=1}^N \left[\int_a^N db f_k(a,b) - \frac{1}{12} \partial_2 f_k(a,a)\right] \label{eq:S-k-alb}
\end{align}
where $\partial_2 f_k(a,b)$ denotes the derivative of the second argument of $f_k(a,b)$ and the $\partial_2 f_k(a,N)$ and $f_k(a,N)$ terms have been dropped with the foresight that they will not contribute at order $O(N)$. The $\partial_2 f_k(a,a)$ term will in fact also not contribute at order $O(N)$, but yields an important singularity at order $O(N^{\alpha})$, so we will bring it along the manipulations.

We first consider the integral term. This inner integral is what ultimately induces the localization to the diagonal where $a = b$. We first change variables of the integral to $x$, and so pick up our density $\rho(x)$: \begin{align}
    \int_a^N db f_k(a,b) &= \int_{x_a}^{x_N} dx \frac{db}{dx} e^{ik(h_1(x_a) + h_2(x)) - kN^\alpha|x_a - x|}\\
    &= N \int_{x_a}^{x_N} dx \rho(x) e^{ik(h_1(x_a) + h_2(x)) - kN^\alpha(x - x_a)}
\end{align}
since we have restricted to the region $a < b$, we have also been able to replace the absolute value. At this stage, the key trick for localization is to Taylor expand the non-exponential portion of the integrand around $x = x_a$. After much manipulation, this yields \begin{align}
    \int_a^N db f_k(a,b) &=N e^{ik h_1(x_a)}\int_{x_a}^{x_N} dx \sum_{l=0}^\infty \partial_k^l \left[\rho(x) e^{ik h_2(x)}\right]\Big|_{x = x_a} \frac{(x - x_a)^l}{l!} e^{- kN^\alpha(x - x_a)}\\
    &= N e^{ik h_1(x_a)}\sum_{l=0}^\infty \partial_k^l \left[\rho(x) e^{ik h_2(x)}\right]\Big|_{x = x_a} \int_{x_a}^{x_N} dx \frac{(x - x_a)^l}{l!} e^{- kN^\alpha(x - x_a)}\\
    &= N e^{ik h_1(x_a)}\sum_{l=0}^\infty\partial_k^l \left[\rho(x) e^{ik h_2(x)}\right]\Big|_{x = x_a} \frac{1}{(-N^\alpha)^l l!}\partial_k^l \left[\int_{x_a}^{x_N} dx e^{- kN^\alpha(x - x_a)}\right]\\
    &= N e^{ik h_1(x_a)}\sum_{l=0}^\infty\partial_k^l \left[\rho(x) e^{ik h_2(x)}\right]\Big|_{x = x_a} \frac{1}{(-N^\alpha)^l l!}\partial_k^l \left[\frac{1 - e^{-kN^\alpha(x_N - x_a)}}{k N^\alpha}\right]
\end{align}
At this point, we can notice that each term in $l$ comes with a suppression of $N^\alpha$. If we are only interested in terms which contribute to $O(N)$, and for many theories we consider $\alpha = \frac{1}{2}$, only the $l = 0$ and $l = 1$ terms of this series will be relevant. Writing those out, we have \begin{align}
     \int_a^N db f_k(a,b) &= \frac{N^{1 - \alpha}}{k} \rho(x_a) e^{ik(h_1(x_a) + h_2(x_a))} \left(1 - e^{-k N^\alpha (x_N - x_a)}\right) \\
     &+ \frac{N^{1 - 2 \alpha}}{k^2} e^{ik(h_1(x_a) + h_2(x_a))}\left(\rho'(x_a) + \rho(x) i k h_2'(x_a)\right)\partial_k \left[\frac{1 - e^{-kN^\alpha(x_N - x_a)}}{k N^\alpha}\right]\label{eq:int-fk}
\end{align}
At this stage, we note that the other half of the sum $S_{k,a>b}$ can be found from $S_{k,a<b}$ under the replacement $x_b \mapsto x_{N - b}$. Because of this, the same manipulation will flip the sign of first order derivatives like $\rho'(x_a)$ and $h_2'(x_a)$ that we encounter. Since we will be adding the result for $S_{k,a>b}$, these terms will then cancel and we can drop the $l = 1$ contribution (\ref{eq:int-fk}) altogether.  

We can now return our attention to (\ref{eq:S-k-alb}) and treat the outer sum over $a$ with the form (\ref{eq:EM-Same}) of the Euler-Maclaurin formula. In this case, however, all of the boundary correction terms will contribute at order below $O(N)$, so we simply replace the sum with an integral: \begin{align}
    &S_{k,a<b}=\int_1^N da \left[\frac{N^{1 - \alpha}}{k} \rho(x_a) e^{ik(h_1(x_a) + h_2(x_a))} \left(1 - e^{-k N^\alpha (x_N - x_a)}\right) - \frac{1}{12} \partial_2 f_k(a,a)\right]\\
    &= \int_{x_1}^{x_N} dx N \rho(x)\left[\frac{N^{1 - \alpha}}{k} \rho(x) e^{ik(h_1(x) + h_2(x))} \left(1 - e^{-k N^\alpha (x_N - x)}\right) - \frac{1}{12} \partial_2 f_k(a,a)\right]\\
    &= \frac{N^{2 - \alpha}}{k}\int dx \rho(x)^2 e^{ik(h_1(x)+h_2(x))} - \frac{N^{2 - \alpha}}{k}\int_{x_1}^{x_N} dx \rho(x)^2 e^{i k (h_1(x) + h_2(x))} e^{-kN^\alpha(x_N - x)} \\
    &- \frac{N}{12}\int_{x_1}^{x_n}dx \rho(x) \partial_2 f_k(a,a)
\end{align}

We can apply a familiar localization trick to the second integral, this time expanding about $x = x_2$. In this case, only the $l = 0$ term contributes at order $O(N)$: \begin{align}
    &\int_{x_1}^{x_N} dx \rho(x) e^{i k (h_1(x) + h_2(x))} e^{-kN^\alpha(x_N - x)}\\
    &= \int_{x_1}^{x_N} dx \sum_{l=0}^\infty \frac{(x - x_N)^l}{l!}\partial_x^l \left[\rho(x)^2 e^{i k (h_1(x) + h_2(x))}\right]\Big|_{x = x_N} e^{-kN^\alpha(x_N - x)}\\
    &= \rho(x_N)^2 e^{ik(h_1(x_N) + h_2(x_N))} \int_{x_1}^{x_N} e^{-k N^\alpha (x_N - x)}\\
    &= \frac{1}{kN^\alpha}\rho(x_N)^2 e^{ik(h_1(x_N) + h_2(x_N))}+O(e^{-N})
\end{align}

We may also compute that \begin{align}
    \partial_2 f_k(a,a) &= \frac{dx}{db} \partial_x \left[e^{ik(h_1(x_a) + h_2(x))-kN^\alpha(x - x_a)}\right]\Big|_{x = x_a}\\
    &= \frac{1}{\rho(x_a) N}\left[e^{ik(h_1(x_a) + h_2(x_a))} k N^\alpha + i k h_2'(x_a) e^{ik(h_1(x_a)+h_2(x_a))}\right]
\end{align}
The second term here will not contribute at leading order. If we add the analogous result for the other half $S_{k,a>b}$, we may assemble the large $N$ expansion: \begin{align}
    \sum_{a=1}^N \sum_{b=1}^N f_k(a,b) &= \frac{2 N^{2 - \alpha}}{k} \int dx \rho(x)^2 e^{ik(h_1(x) + h_2(x))} - \frac{kN^\alpha}{6} \int dx e^{ik(h_1(x) + h_2(x))}\\
    &- \frac{N^{2 - 2 \alpha}}{k^2} \left(\rho(x_1)^2e^{ik(h_1(x_1)+h_2(x_1))} + \rho(x_N)^2e^{ik(h_1(x_N)+h_2(x_N))}\right)
\end{align}

If we then perform the sum over $k$, we obtain the final result: \begin{align}
    &\sum_{a=1}^N \sum_{b=1}^N \text{Li}_1\left(e^{i(h_1(x_a) + h_2(x_b)) - N^\alpha|x_{ab}|}\right) = \sum_{k=0}^\infty \frac{1}{k} \sum_{a=1}^N \sum_{b=1}^N f_k(a,b)\\
    &= \sum_{k=0}^\infty \Big[2 N^{2 - \alpha} \int dx \rho(x)^2 \frac{1}{k^2}e^{ik(h_1(x) + h_2(x))} - \frac{N^\alpha}{6} \int dx e^{ik(h_1(x) + h_2(x))}\\
    &- N^{2 - 2 \alpha} \left(\rho(x_1)^2\frac{1}{k^3}e^{ik(h_1(x_1)+h_2(x_1))} + \rho(x_N)^2\frac{1}{k^3}e^{ik(h_1(x_N)+h_2(x_N))}\right)\Big]\\
    &= 2 N^{2 - \alpha} \int dx \rho(x)^2 \text{Li}_2\left(e^{i(h_1(x)+h_2(x))}\right) - \frac{N^{\alpha}}{6} \int dx \text{Li}_0\left(e^{i(h_1(x)+h_2(x))}\right) \\
    &- N^{2 - 2\alpha}\left(\rho(x_1)^2\text{Li}_3 \left(e^{i(h_1(x_1)+h_2(x_1))}\right)+\rho(x_N)^2\text{Li}_3 \left(e^{i(h_1(x_N)+h_2(x_N))}\right)\right).
\end{align}
Since \begin{align}
    \text{Li}_0(e^{iz}) \sim \frac{i}{z}
\end{align}
We see that there is indeed a singularity at $h_1(x) + h_2(x) = 0$ at order $O(N^\alpha)$. 

\subsection{Gauge node term}
The large $N$ expansion of the gauge node term requires a slightly different approach than the method above aimed at the chiral contributions. The sum in question is now \begin{align}
    \sum_{a \not= b} \text{Li}_1 \left(e^{iy_{ab}-N^\alpha|x_{ab}|}\right) \label{eq:Li-1-DSNE}
\end{align} as well as the same sum with $\tilde{y}$ instead of $y$. 

With this sum it is not possible to simply use the result of (\ref{eq:Li-1-DS}) and subtract off the $a = b$ terms since both are singular. Instead, it is necessary to repeat the process of applying the Euler-Maclaurin formula except now choosing the form (\ref{eq:EM-Same}) to represent the inner sum. This ensures that the integral does not run all the way to the diagonal where it will diverge. 

If this (equally long) path is followed, then the same leading contribution as in the chiral case is obtained. The subleading in $N$ terms do then exhibit divergences, but they are in fact divergences in $N$ which yield $N \log N$ terms. Due to the lack of singularities in $\delta y(x)$, these effects should not be included in $\mathcal{L}_{\text{ABJM}}^{(sub)}$. Another way to understand that the final result does not depend upon $y$ or $\tilde{y}$ is that the double sums naturally localize to the $a = b$ diagonal, where $y_{ab} \equiv \tilde{y}_{ab} \equiv 0$, so there is no variation in those variables available to even realize a singularity. 

Ultimately, the relevant contribution to the Lagrangian is simply the leading term: \begin{align}
    &\sum_{a \not= b} \left[ \text{Li}_1 \left(e^{iy_{ab}-N^\alpha|x_{ab}|}\right) + \text{Li}_1 \left(e^{i\tilde{y}_{ab}-N^\alpha|x_{ab}|}\right)\right] \nonumber  \\
    &= \int dx 2 \rho(x)^2 N^{2-\alpha}\text{Li}_2 \left(1\right) = \frac{\pi^2}{3} \int dx \rho(x)^2
\end{align}

\subsection{Numerical Checks}
To check that these large $N$ manipulations have indeed yielded the correct results, we can perform some numerical experiments. While we cannot perform exhaustive checks, we have considered multiple examples, some of which we report in this section. As $N \rightarrow \infty$, we must define choices of $x_a$ for $a = 1,...,N$ which converge to a fixed distribution $\rho(x)$. We consider a piecewise-linear example of $\rho(x)$, keeping with the form of the distribution that we find for ABJM. Specifically, we choose \begin{align}
    \rho(x) = \begin{cases}
        0 & x \in (-\infty, -1) \cup (1,\infty) \\
        1 + x & x \in [-1,0]\\
        1 - x & x \in (0,1]
    \end{cases}
\end{align}

And our choices of $x_a$ which will converge to this distribution are \begin{align}
    x_a = \begin{cases}
        -1 + \sqrt{\frac{2 a - 1}{N}} & a \leq \frac{N}{2} \\
        1 - \sqrt{\frac{2 (N-a) - 1}{N}} & a > \frac{N}{2}
    \end{cases} \label{eq:xa-choice}
\end{align}
To check (\ref{eq:Li-1-DS}), we will choose $h_1(x) = x^2$, $h_2(x) = -x$, and $\alpha = \frac{1}{2}$. With this choice, we would like to check the result that \begin{align}
    \sum_{a,b} \text{Li}_1 \left(e^{i(x_a^2 - x_b)- N^{1/2} |x_{ab}|}\right) &= \left[\int \rho(x)^2 \text{Li}_2\left(e^{i(x^2 - x)}\right) dx\right] 2  N^{3/2} \\
    &= 2(0.8562+0.057i)N^{3/2} + o(N) \label{eq:explicit-Check}
\end{align}

\begin{figure}[t]
\centering
\includegraphics[width=0.7\textwidth]{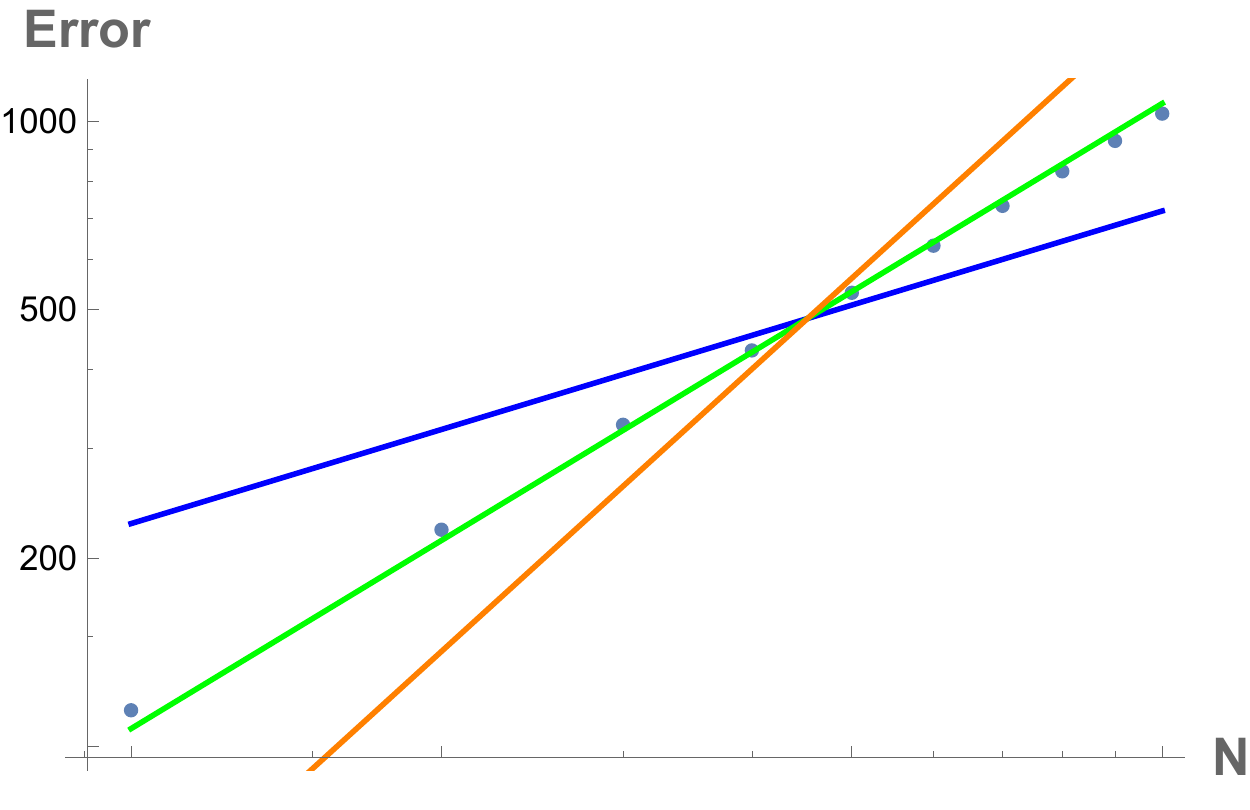}
\caption{Plot of the error of our claimed large-$N$ expression for (\ref{eq:explicit-Check}), evaluated at $N \in \{200,...,1000\}$. The error of only the real part is plotted to avoid branch-cut issues. The blue line indicates $O(N^{1/2})$ growth, green is $O(N)$, and orange is $O(N^{3/2})$. The error appears to be of order $O(N^{1/2})$.}\label{fig:error-N}
\end{figure}

To check this, we can explicitly evaluate the sum and compare against our claim. As shown in Fig.~\ref{fig:error-N}, an error of scale $O(N^{1/2})$ is observed. This is evidence that we have properly recovered the $O(N^{3/2})$ and $O(N)$ behavior. 

To demonstrate the validity of the $O(N^{2 - 2 \alpha})$ term in (\ref{eq:Li-1-DS}), we must consider a distribution which does not go to zero at its endpoints. One choice for this could be the uniform distribution on $x \in [0,1]$. A corresponding choice of $x_a$ that converges to this distribution could then be $x_a = \frac{a}{N}$. We will also consider constant $h_1(x) = h_2(x) = \frac{y}{2}$ and again $\alpha = \frac{1}{2}$. In this case, our claimed large $N$ expansion would be \begin{align}
    \sum_{a,b} \text{Li}_1 \left(e^{iy- N^{1/2} |\frac{a}{N} - \frac{b}{N}|}\right) &= 2 N^{3/2}\int dx \rho(x)^2 \text{Li}_2\left(e^{i y}\right) - \frac{N^{1/2}}{6}\int dx \text{Li}_0\left(e^{i y}\right)\\
    &- N \left(\rho(x_1)^2 \text{Li}_3\left(e^{i y }\right) + \rho(x_N)^2 \text{Li}_3\left(e^{i y }\right)\right)\\
    &= 2 N^{3/2}\text{Li}_2\left(e^{i y}\right) - \frac{N^{1/2}}{6} \text{Li}_0\left(e^{i y}\right)-2N \text{Li}_3\left(e^{i y }\right) + o(N) \label{eq:explicit-Check-2}
\end{align}

\begin{figure}[t]
\centering
\includegraphics[width=0.7\textwidth]{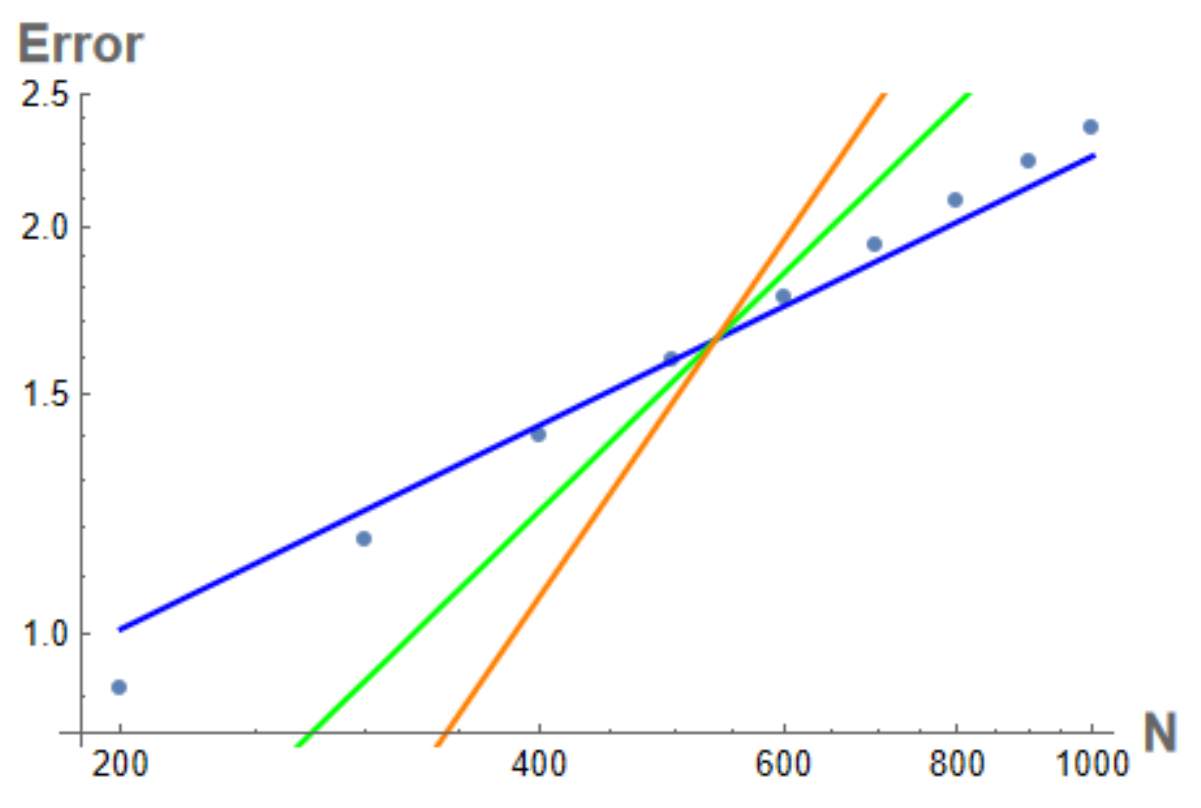}
\caption{Plot of the error of our claimed large-$N$ expression for (\ref{eq:explicit-Check-2}), evaluated at $N \in \{200,...,1000\}$ and colored as in Fig.~\ref{fig:error-N}. The error appears to again be of order $O(N^{1/2})$.}\label{fig:error-N-2}
\end{figure}

In Figure~\ref{fig:error-N-2} this claim is tested for $y = 0.2$, and again an $O(N^{1/2})$ error is observed, implying that both the $O(N^{3/2})$ and $O(N)$ behavior of the term have been correctly described. 

\bibliographystyle{JHEP}
\bibliography{BHLocalization}

\providecommand{\href}[2]{#2}\begingroup\raggedright\begin{thebibliography}{10}

\bibitem{Choi:2019zpz}
S.~Choi, C.~Hwang and S.~Kim, \emph{{Quantum vortices, M2-branes and black
  holes}},  \href{https://arxiv.org/abs/1908.02470}{{\ttfamily 1908.02470}}.

\bibitem{Nian:2019pxj}
J.~Nian and L.A.~Pando~Zayas, \emph{{Microscopic entropy of rotating
  electrically charged AdS$_{4}$ black holes from field theory localization}},
  \href{https://doi.org/10.1007/JHEP03(2020)081}{\emph{JHEP} {\bfseries 03}
  (2020) 081} [\href{https://arxiv.org/abs/1909.07943}{{\ttfamily
  1909.07943}}].

\bibitem{Cabo-Bizet:2018ehj}
A.~Cabo-Bizet, D.~Cassani, D.~Martelli and S.~Murthy, \emph{{Microscopic origin
  of the Bekenstein-Hawking entropy of supersymmetric AdS$_{\bf 5}$ black
  holes}},  \href{https://arxiv.org/abs/1810.11442}{{\ttfamily 1810.11442}}.

\bibitem{Choi:2018hmj}
S.~Choi, J.~Kim, S.~Kim and J.~Nahmgoong, \emph{{Large AdS black holes from
  QFT}},  \href{https://arxiv.org/abs/1810.12067}{{\ttfamily 1810.12067}}.

\bibitem{Benini:2018ywd}
F.~Benini and P.~Milan, \emph{{Black Holes in 4D $\mathcal{N}$=4
  Super-Yang-Mills Field Theory}},
  \href{https://doi.org/10.1103/PhysRevX.10.021037}{\emph{Phys. Rev. X}
  {\bfseries 10} (2020) 021037}
  [\href{https://arxiv.org/abs/1812.09613}{{\ttfamily 1812.09613}}].

\bibitem{Liu:2017vbl}
J.T.~Liu, L.A.~Pando~Zayas, V.~Rathee and W.~Zhao, \emph{{One-Loop Test of
  Quantum Black Holes in anti–de Sitter Space}},
  \href{https://doi.org/10.1103/PhysRevLett.120.221602}{\emph{Phys. Rev. Lett.}
  {\bfseries 120} (2018) 221602}
  [\href{https://arxiv.org/abs/1711.01076}{{\ttfamily 1711.01076}}].

\bibitem{Gang:2019uay}
D.~Gang, N.~Kim and L.A.~Pando~Zayas, \emph{{Precision Microstate Counting for
  the Entropy of Wrapped M5-branes}},
  \href{https://doi.org/10.1007/JHEP03(2020)164}{\emph{JHEP} {\bfseries 03}
  (2020) 164} [\href{https://arxiv.org/abs/1905.01559}{{\ttfamily
  1905.01559}}].

\bibitem{Benini:2019dyp}
F.~Benini, D.~Gang and L.A.~Pando~Zayas, \emph{{Rotating Black Hole Entropy
  from M5 Branes}}, \href{https://doi.org/10.1007/JHEP03(2020)057}{\emph{JHEP}
  {\bfseries 03} (2020) 057}
  [\href{https://arxiv.org/abs/1909.11612}{{\ttfamily 1909.11612}}].

\bibitem{PandoZayas:2020iqr}
L.A.~Pando~Zayas and Y.~Xin, \emph{{Universal logarithmic behavior in
  microstate counting and the dual one-loop entropy of $AdS_4$ black holes}},
  \href{https://doi.org/10.1103/PhysRevD.103.026003}{\emph{Phys. Rev. D}
  {\bfseries 103} (2021) 026003}
  [\href{https://arxiv.org/abs/2008.03239}{{\ttfamily 2008.03239}}].

\bibitem{Bobev:2020egg}
N.~Bobev, A.M.~Charles, K.~Hristov and V.~Reys, \emph{{The Unreasonable
  Effectiveness of Higher-Derivative Supergravity in AdS$_4$ Holography}},
  \href{https://doi.org/10.1103/PhysRevLett.125.131601}{\emph{Phys. Rev. Lett.}
  {\bfseries 125} (2020) 131601}
  [\href{https://arxiv.org/abs/2006.09390}{{\ttfamily 2006.09390}}].

\bibitem{Bobev:2020zov}
N.~Bobev, A.M.~Charles, D.~Gang, K.~Hristov and V.~Reys,
  \emph{{Higher-derivative supergravity, wrapped M5-branes, and theories of
  class $ \mathrm{\mathcal{R}} $}},
  \href{https://doi.org/10.1007/JHEP04(2021)058}{\emph{JHEP} {\bfseries 04}
  (2021) 058} [\href{https://arxiv.org/abs/2011.05971}{{\ttfamily
  2011.05971}}].

\bibitem{Bobev:2021oku}
N.~Bobev, A.M.~Charles, K.~Hristov and V.~Reys, \emph{{Higher-derivative
  supergravity, AdS$_{4}$ holography, and black holes}},
  \href{https://doi.org/10.1007/JHEP08(2021)173}{\emph{JHEP} {\bfseries 08}
  (2021) 173} [\href{https://arxiv.org/abs/2106.04581}{{\ttfamily
  2106.04581}}].

\bibitem{Ghosh:2020rwf}
J.K.~Ghosh and L.A.~Pando~Zayas, \emph{{Comments on Sen's Classical Entropy
  Function for Static and Rotating AdS$_4$ Black Holes}},
  \href{https://arxiv.org/abs/2009.11147}{{\ttfamily 2009.11147}}.

\bibitem{Liu:2022sew}
J.T.~Liu and R.J.~Saskowski, \emph{{Four-derivative corrections to minimal
  gauged supergravity in five dimensions}},
  \href{https://doi.org/10.1007/JHEP05(2022)171}{\emph{JHEP} {\bfseries 05}
  (2022) 171} [\href{https://arxiv.org/abs/2201.04690}{{\ttfamily
  2201.04690}}].

\bibitem{Bobev:2022bjm}
N.~Bobev, V.~Dimitrov, V.~Reys and A.~Vekemans, \emph{{Higher-Derivative
  Corrections and AdS$_5$ Black Holes}},
  \href{https://arxiv.org/abs/2207.10671}{{\ttfamily 2207.10671}}.

\bibitem{Cassani:2022lrk}
D.~Cassani, A.~Ruip\'erez and E.~Turetta, \emph{{Corrections to AdS$_5$ Black
  Hole Thermodynamics from Higher-Derivative Supergravity}},
  \href{https://arxiv.org/abs/2208.01007}{{\ttfamily 2208.01007}}.

\bibitem{GonzalezLezcano:2020yeb}
A.~Gonz\'alez~Lezcano, J.~Hong, J.T.~Liu and L.A.~Pando~Zayas,
  \emph{{Sub-leading Structures in Superconformal Indices: Subdominant Saddles
  and Logarithmic Contributions}},
  \href{https://doi.org/10.1007/JHEP01(2021)001}{\emph{JHEP} {\bfseries 01}
  (2021) 001} [\href{https://arxiv.org/abs/2007.12604}{{\ttfamily
  2007.12604}}].

\bibitem{Lezcano:2021qbj}
A.G.~Lezcano, J.~Hong, J.T.~Liu and L.A.~Pando~Zayas, \emph{{The Bethe-Ansatz
  approach to the $ \mathcal{N} $ = 4 superconformal index at finite rank}},
  \href{https://doi.org/10.1007/JHEP06(2021)126}{\emph{JHEP} {\bfseries 06}
  (2021) 126} [\href{https://arxiv.org/abs/2101.12233}{{\ttfamily
  2101.12233}}].

\bibitem{Amariti:2020jyx}
A.~Amariti, M.~Fazzi and A.~Segati, \emph{{The SCI of $ \mathcal{N} $ = 4
  USp(2N$_{c}$) and SO(N$_{c}$) SYM as a matrix integral}},
  \href{https://doi.org/10.1007/JHEP06(2021)132}{\emph{JHEP} {\bfseries 06}
  (2021) 132} [\href{https://arxiv.org/abs/2012.15208}{{\ttfamily
  2012.15208}}].

\bibitem{Amariti:2021ubd}
A.~Amariti, M.~Fazzi and A.~Segati, \emph{{Expanding on the Cardy-like limit of
  the SCI of 4d $ \mathcal{N} $ = 1 ABCD SCFTs}},
  \href{https://doi.org/10.1007/JHEP07(2021)141}{\emph{JHEP} {\bfseries 07}
  (2021) 141} [\href{https://arxiv.org/abs/2103.15853}{{\ttfamily
  2103.15853}}].

\bibitem{Cassani:2021fyv}
D.~Cassani and Z.~Komargodski, \emph{{EFT and the SUSY Index on the 2nd
  Sheet}}, \href{https://doi.org/10.21468/SciPostPhys.11.1.004}{\emph{SciPost
  Phys.} {\bfseries 11} (2021) 004}
  [\href{https://arxiv.org/abs/2104.01464}{{\ttfamily 2104.01464}}].

\bibitem{ArabiArdehali:2021nsx}
A.~Arabi~Ardehali and S.~Murthy, \emph{{The 4d superconformal index near roots
  of unity and 3d Chern-Simons theory}},
  \href{https://arxiv.org/abs/2104.02051}{{\ttfamily 2104.02051}}.

\bibitem{David:2020ems}
M.~David, J.~Nian and L.A.~Pando~Zayas, \emph{{Gravitational Cardy Limit and
  AdS Black Hole Entropy}},
  \href{https://doi.org/10.1007/JHEP11(2020)041}{\emph{JHEP} {\bfseries 11}
  (2020) 041} [\href{https://arxiv.org/abs/2005.10251}{{\ttfamily
  2005.10251}}].

\bibitem{Bardeen:1999px}
J.M.~Bardeen and G.T.~Horowitz, \emph{{The Extreme Kerr throat geometry: A
  Vacuum analog of AdS(2) x S**2}},
  \href{https://doi.org/10.1103/PhysRevD.60.104030}{\emph{Phys. Rev. D}
  {\bfseries 60} (1999) 104030}
  [\href{https://arxiv.org/abs/hep-th/9905099}{{\ttfamily hep-th/9905099}}].

\bibitem{Guica:2008mu}
M.~Guica, T.~Hartman, W.~Song and A.~Strominger, \emph{{The Kerr/CFT
  Correspondence}},
  \href{https://doi.org/10.1103/PhysRevD.80.124008}{\emph{Phys. Rev. D}
  {\bfseries 80} (2009) 124008}
  [\href{https://arxiv.org/abs/0809.4266}{{\ttfamily 0809.4266}}].

\bibitem{Lu:2008jk}
H.~Lu, J.~Mei and C.N.~Pope, \emph{{Kerr/CFT Correspondence in Diverse
  Dimensions}},
  \href{https://doi.org/10.1088/1126-6708/2009/04/054}{\emph{JHEP} {\bfseries
  04} (2009) 054} [\href{https://arxiv.org/abs/0811.2225}{{\ttfamily
  0811.2225}}].

\bibitem{David:2021qaa}
M.~David, A.~Lezcano~Gonz\'alez, J.~Nian and L.A.~Pando~Zayas,
  \emph{{Logarithmic corrections to the entropy of rotating black holes and
  black strings in AdS$_{5}$}},
  \href{https://doi.org/10.1007/JHEP04(2022)160}{\emph{JHEP} {\bfseries 04}
  (2022) 160} [\href{https://arxiv.org/abs/2106.09730}{{\ttfamily
  2106.09730}}].

\bibitem{David:2021eoq}
M.~David, V.~Godet, Z.~Liu and L.A.~Pando~Zayas, \emph{{Non-topological
  logarithmic corrections in minimal gauged supergravity}},
  \href{https://doi.org/10.1007/JHEP08(2022)043}{\emph{JHEP} {\bfseries 08}
  (2022) 043} [\href{https://arxiv.org/abs/2112.09444}{{\ttfamily
  2112.09444}}].

\bibitem{Romelsberger:2005eg}
C.~Romelsberger, \emph{{Counting chiral primaries in N = 1, d=4 superconformal
  field theories}},
  \href{https://doi.org/10.1016/j.nuclphysb.2006.03.037}{\emph{Nucl. Phys. B}
  {\bfseries 747} (2006) 329}
  [\href{https://arxiv.org/abs/hep-th/0510060}{{\ttfamily hep-th/0510060}}].

\bibitem{Kinney:2005ej}
J.~Kinney, J.M.~Maldacena, S.~Minwalla and S.~Raju, \emph{{An Index for 4
  dimensional super conformal theories}},
  \href{https://doi.org/10.1007/s00220-007-0258-7}{\emph{Commun. Math. Phys.}
  {\bfseries 275} (2007) 209}
  [\href{https://arxiv.org/abs/hep-th/0510251}{{\ttfamily hep-th/0510251}}].

\bibitem{Bhattacharya:2008zy}
J.~Bhattacharya, S.~Bhattacharyya, S.~Minwalla and S.~Raju, \emph{{Indices for
  Superconformal Field Theories in 3,5 and 6 Dimensions}},
  \href{https://doi.org/10.1088/1126-6708/2008/02/064}{\emph{JHEP} {\bfseries
  02} (2008) 064} [\href{https://arxiv.org/abs/0801.1435}{{\ttfamily
  0801.1435}}].

\bibitem{Kim:2009wb}
S.~Kim, \emph{{The Complete superconformal index for N=6 Chern-Simons theory}},
  \href{https://doi.org/10.1016/j.nuclphysb.2009.06.025}{\emph{Nucl. Phys. B}
  {\bfseries 821} (2009) 241}
  [\href{https://arxiv.org/abs/0903.4172}{{\ttfamily 0903.4172}}].

\bibitem{Choi:2019dfu}
S.~Choi and C.~Hwang, \emph{{Universal 3d Cardy Block and Black Hole Entropy}},
  \href{https://doi.org/10.1007/JHEP03(2020)068}{\emph{JHEP} {\bfseries 03}
  (2020) 068} [\href{https://arxiv.org/abs/1911.01448}{{\ttfamily
  1911.01448}}].

\bibitem{Imamura:2011wg}
Y.~Imamura and D.~Yokoyama, \emph{{N=2 supersymmetric theories on squashed
  three-sphere}}, {\emph{Phys.Rev.} {\bfseries D85} (2012) 025015}
  [\href{https://arxiv.org/abs/1109.4734}{{\ttfamily 1109.4734}}].

\bibitem{Aharony:2008ug}
O.~Aharony, O.~Bergman, D.L.~Jafferis and J.~Maldacena, \emph{{N=6
  superconformal Chern-Simons-matter theories, M2-branes and their gravity
  duals}}, \href{https://doi.org/10.1088/1126-6708/2008/10/091}{\emph{JHEP}
  {\bfseries 10} (2008) 091} [\href{https://arxiv.org/abs/0806.1218}{{\ttfamily
  0806.1218}}].

\bibitem{Benini:2015eyy}
F.~Benini, K.~Hristov and A.~Zaffaroni, \emph{{Black hole microstates in
  AdS$_{4}$ from supersymmetric localization}},
  \href{https://doi.org/10.1007/JHEP05(2016)054}{\emph{JHEP} {\bfseries 05}
  (2016) 054} [\href{https://arxiv.org/abs/1511.04085}{{\ttfamily
  1511.04085}}].

\bibitem{Liu:2017vll}
J.T.~Liu, L.A.~Pando~Zayas, V.~Rathee and W.~Zhao, \emph{{Toward Microstate
  Counting Beyond Large N in Localization and the Dual One-loop Quantum
  Supergravity}}, \href{https://doi.org/10.1007/JHEP01(2018)026}{\emph{JHEP}
  {\bfseries 01} (2018) 026}
  [\href{https://arxiv.org/abs/1707.04197}{{\ttfamily 1707.04197}}].

\bibitem{Hosseini:2022vho}
S.M.~Hosseini and A.~Zaffaroni, \emph{{The large $N$ limit of topologically
  twisted indices: a direct approach}},
  \href{https://arxiv.org/abs/2209.09274}{{\ttfamily 2209.09274}}.

\bibitem{Beem:2012mb}
C.~Beem, T.~Dimofte and S.~Pasquetti, \emph{{Holomorphic Blocks in Three
  Dimensions}}, \href{https://doi.org/10.1007/JHEP12(2014)177}{\emph{JHEP}
  {\bfseries 12} (2014) 177} [\href{https://arxiv.org/abs/1211.1986}{{\ttfamily
  1211.1986}}].

\bibitem{Bhattacharyya:2007vs}
S.~Bhattacharyya, S.~Lahiri, R.~Loganayagam and S.~Minwalla, \emph{{Large
  rotating AdS black holes from fluid mechanics}},
  \href{https://doi.org/10.1088/1126-6708/2008/09/054}{\emph{JHEP} {\bfseries
  09} (2008) 054} [\href{https://arxiv.org/abs/0708.1770}{{\ttfamily
  0708.1770}}].

\bibitem{Shaghoulian:2015kta}
E.~Shaghoulian, \emph{{Modular forms and a generalized Cardy formula in higher
  dimensions}}, \href{https://doi.org/10.1103/PhysRevD.93.126005}{\emph{Phys.
  Rev. D} {\bfseries 93} (2016) 126005}
  [\href{https://arxiv.org/abs/1508.02728}{{\ttfamily 1508.02728}}].

\bibitem{Chong:2005hr}
Z.W.~Chong, M.~Cvetic, H.~Lu and C.N.~Pope, \emph{{General non-extremal
  rotating black holes in minimal five-dimensional gauged supergravity}},
  \href{https://doi.org/10.1103/PhysRevLett.95.161301}{\emph{Phys. Rev. Lett.}
  {\bfseries 95} (2005) 161301}
  [\href{https://arxiv.org/abs/hep-th/0506029}{{\ttfamily hep-th/0506029}}].

\bibitem{Cvetic:2005zi}
M.~Cvetic, G.W.~Gibbons, H.~Lu and C.N.~Pope, \emph{{Rotating black holes in
  gauged supergravities: Thermodynamics, supersymmetric limits, topological
  solitons and time machines}},
  \href{https://arxiv.org/abs/hep-th/0504080}{{\ttfamily hep-th/0504080}}.

\bibitem{Larsen:2021wnu}
F.~Larsen and S.~Lee, \emph{{Microscopic entropy of AdS$_{3}$ black holes
  revisited}}, \href{https://doi.org/10.1007/JHEP07(2021)038}{\emph{JHEP}
  {\bfseries 07} (2021) 038}
  [\href{https://arxiv.org/abs/2101.08497}{{\ttfamily 2101.08497}}].

\bibitem{Agarwal:2020zwm}
P.~Agarwal, S.~Choi, J.~Kim, S.~Kim and J.~Nahmgoong, \emph{{AdS black holes
  and finite N indices}},
  \href{https://doi.org/10.1103/PhysRevD.103.126006}{\emph{Phys. Rev. D}
  {\bfseries 103} (2021) 126006}
  [\href{https://arxiv.org/abs/2005.11240}{{\ttfamily 2005.11240}}].

\bibitem{Murthy:2022tbj}
S.~Murthy, \emph{{Growth of the $\frac {1} {16}$-BPS index in 4d $N=4$
  supersymmetric Yang-Mills theory}},
  \href{https://doi.org/10.1103/PhysRevD.105.L021903}{\emph{Phys. Rev. D}
  {\bfseries 105} (2022) L021903}.

\bibitem{Choi:2018fdc}
S.~Choi, C.~Hwang, S.~Kim and J.~Nahmgoong, \emph{{Entropy Functions of BPS
  Black Holes in AdS$_{4}$ and AdS$_{6}$}},
  \href{https://doi.org/10.3938/jkps.76.101}{\emph{J. Korean Phys. Soc.}
  {\bfseries 76} (2020) 101}
  [\href{https://arxiv.org/abs/1811.02158}{{\ttfamily 1811.02158}}].

\bibitem{Cassani:2019mms}
D.~Cassani and L.~Papini, \emph{{The BPS limit of rotating AdS black hole
  thermodynamics}}, \href{https://doi.org/10.1007/JHEP09(2019)079}{\emph{JHEP}
  {\bfseries 09} (2019) 079}
  [\href{https://arxiv.org/abs/1906.10148}{{\ttfamily 1906.10148}}].

\bibitem{Bobev:2022jte}
N.~Bobev, J.~Hong and V.~Reys, \emph{{Large N Partition Functions, Holography,
  and Black Holes}},
  \href{https://doi.org/10.1103/PhysRevLett.129.041602}{\emph{Phys. Rev. Lett.}
  {\bfseries 129} (2022) 041602}
  [\href{https://arxiv.org/abs/2203.14981}{{\ttfamily 2203.14981}}].

\bibitem{DiPietro:2014bca}
L.~Di~Pietro and Z.~Komargodski, \emph{{Cardy formulae for SUSY theories in $d
  =$ 4 and $d =$ 6}},
  \href{https://doi.org/10.1007/JHEP12(2014)031}{\emph{JHEP} {\bfseries 12}
  (2014) 031} [\href{https://arxiv.org/abs/1407.6061}{{\ttfamily 1407.6061}}].

\bibitem{Hristov:2022lcw}
K.~Hristov, \emph{{ABJM at finite $N$ via 4d supergravity}},
  \href{https://arxiv.org/abs/2204.02992}{{\ttfamily 2204.02992}}.

\bibitem{Garoufalidis:2018qds}
S.~Garoufalidis and D.~Zagier, \emph{{Asymptotics of Nahm sums at roots of
  unity}},  \href{https://arxiv.org/abs/1812.07690}{{\ttfamily 1812.07690}}.

\end{thebibliography}\endgroup
\end{document}